\documentclass[aps,notitlepage,prb,10pt]{revtex4-2}
\usepackage[caption=false]{subfig}
\usepackage{makeidx}
\usepackage{eurosym}
\usepackage{graphicx}
\usepackage{graphics}
\usepackage{amssymb}
\usepackage{amsmath}
\usepackage{color}

\DeclareUnicodeCharacter{03B3}{$\gamma$}

\begin{document}

\title[Test]{Field-Induced Superconductor-Insulator Transition in Disordered 2D Electron systems: The case
of amorphous Indium-Oxide thin films.}
\author{Tsofar Maniv and Vladimir Zhuravlev}
\affiliation{Schulich Faculty of Chemistry, Technion-Israel Institute of
	Technology, Haifa 32000, Israel}
\date{\today }

\begin{abstract}
The phenomenon of field-induced superconductor to insulator transition (SIT)
in disordered 2D electron systems has been a subject of controversy since
its discovery in the early 1990s. Here we present a phenomenological
quantitative theory of this phenomenon which is not based exclusively on the
boson-vortex duality used commonly in the field. Within a new
low-temperature framework of the time-dependent Ginzburg-Landau (TDGL)
functional approach to superconducting fluctuations we propose and develop a
scenario in which bosons of Cooper-pair fluctuations (CPFs) condense and
localize in real-space mesoscopic puddles under increasing magnetic field
due to diminishing stiffness of the fluctuation modes at low temperatures in
a broad range of momentum space. Quantum tunneled CPFs relieving the
condensed mesoscopic puddles, which consequently pair break into fermionic
quasi-particle excitations, dominate the thermally activated inter-puddles
transport. The spatially shrinking puddles of CPFs, embedded in expanding
normal-state regions, upon further increasing field, suppress the
quasi-particle excitation gap and so lead to high-field negative
magneto-resistance (MR). Application to amorphous Indium-Oxide thin films
shows good quantitative agreement with experimental sheet resistance data.
In particular, in agreement with the experiment at low temperatures (i.e.
well below the quantum tunneling pair breaking "temperature"), the sheet
resistance isotherms are predicted to show a single crossing point at a
quantum critical field not far below the MR peak.
\end{abstract}

\maketitle

\section{Introduction}

In an ideal 2D electron system under a parallel magnetic field Zeeman spin
splitting destroys spatially uniform spin-singlet superconducting (SC) order
above some critical field, known as the Clogston Chadrasekhar critical field
(\cite{Clogston1962}, \cite{Chandrasekhar1962}). A first-order phase
transition from the SC state to the paramagnetic normal state occurs at this
critical field, where the gain in free energy associated with a condensate
of Cooper pairs compensates for the Zeeman spin-splitting energy of the
Pauli paramagnetic normal state. Realization of such systems can be found in
ultra thin films of light metals, like Al and Be for which the spin-orbit
(SO) coupling is very small \cite{AdamsRev2012}. Furthermore, in such ultra
thin metallic films disorder scatterings of electrons in the normal state
are very pronounced due to their highly nonuniform (granular) nature so that
the sharp transitions look very much like superconductor to insulator (SIT)
transitions \cite{AdamsPRL94}. Nevertheless, combining significant SO
coupling with strong disorder in 2D electron systems could open new avenues
for field-induced SIT under parallel magnetic field: One may wonder whether
the resulting continuous (second-order) phase transitions to
superconductivity in the presence of the large order-parameter fluctuations,
characterizing 2D electron systems, can lead to field-induced SIT even in
the absence of the boson-vortex duality \cite{Fisher90} used commonly in the
field. The issue has been discussed in a series of papers by Gantmakher et
al. \cite{GantmakherPhysicaB2000}, who reported on parallel magnetic
field-induced SIT in thin amorphous Indium Oxide films \cite%
{GantmakherJETPLett2000}.

Similar situation of field-induced SIT under both parallel and perpendicular
magnetic fields has been reported \cite{Mograbi19} (see also \cite%
{RoutPRL2017},\cite{ManivNatCommun2015}) for the high mobility electron
systems formed in the electron-doped interfaces between two insulating
perovskite oxides---SrTiO$_{3}$ and LaAlO$_{3}$ \cite{Ohtomo04},\cite%
{Caviglia08}, in which superconductivity was reported \cite{RoutPRL2017} to
be correlated with strong SO interaction. Large negative MR, extended well
above the field of maximal resistance, has drawn much attention \cite%
{GantmakherJETP96}, particularly in light of the expected near absence of
negative MR in 2D electron systems under parallel magnetic field \cite%
{Fukuyama81},\cite{Fukuyama1985}, \cite{DiezPRL2015}.

A scenario associating both the observed field-induced SIT and the large
high-field negative MR to localized Cooper-pairing has been presented in Ref.%
\cite{GantmakherPhysicaB2000}, assuming the existence near the SIT of two
complementary groups of electrons, paired (bosons) of density $n_{B}$, and
unpaired fermions with density $n_{F}$ ($2n_{B}+n_{F}=n$, the total
density). The negative MR was due in this scenario to breaking of localized
pairs and transformation of bosons into fermions, the latter having higher
mobility.

In Refs.\cite{MZPRB2021}, \cite{MZJPC2023} we have developed a time
dependent Ginzburg-Landau (TDGL) functional approach to a model of
Cooper-pair fluctuations (CPFs) in disordered 2D electron gas with strong SO
scattering (SOS), in an attempt to account quantitatively for the extensive
sheet resistance data reported in \cite{Mograbi19} for the SrTiO$_{3}$/LaAlO$%
_{3}$ (111) interface. However, a microscopic Gorkov-GL approach, based on
the diagrammatic formalism developed by Larkin and Varlamov \cite{LV05},
which has been extended recently to very low temperatures in a series of
papers \cite{GalitLarkinPRB01}, \cite{Glatzetal2011}, \cite{Lopatinetal05}, 
\cite{Khodas2012}, has provided no support for the observed pronounced MR 
\cite{MZ-arXiv2025}.

In the present paper we have generalized the TDGL functional approach of
Refs.\cite{MZPRB2021}, \cite{MZJPC2023} so that it can be applied to
disordered 2D electron systems created in many host substrates including
both the amorphous Indium Oxide films, as well as the SrTiO$_{3}$/LaAlO$_{3}$
(111) interface. Consistently with Ref.\cite{MZJPC2023}, it is found that,
due to the extreme field-induced softening of the fluctuation modes at very
low temperatures, which enhances the CPFs density beyond saturation in real
space mesoscopic puddles, the postulate of grand canonical ensemble of
electrons underlying the microscopic (Gorkov-GL) theory of superconducting
fluctuations is unsubstantiated. On the other hand, the plausible assumption
of dynamical equilibrium between the bosons of CPFs and the unpaired
electrons within the TDGL functional approach enables us to take into
account the residual normal-state conductivity consistently with the
calculated (Aslamazov-Larkin \cite{AL68}) paraconductivity by introducing
phenomenologically the joint effect of quantum tunneling and pair breaking
that prevent over saturation of CPFs puddles.

Consequently, one finds that the diminishing stiffness of the CPF modes,
which sharply suppresses the paraconductivity under increasing fields, and
the localization of the remaining disordered 2D system of unpaired electrons
conspire to introduce large MR following the low field superconductivity.
Furthermore, our analysis of the amorphous Indium Oxide thin film data has
shown that the large high field negative MR is associated with
field-enhanced thermally activated residual conductance. The emerging
scenario of field-induced SIT may be therefore described, in the spirit of
Ref.\cite{GantmakherPhysicaB2000}, in terms of CPFs bosons, localized and
nearly saturated in real-space mesoscopic puddles by the magnetic field, in
dynamical equilibrium with the remaining system of unpaired electrons
confined in a spatially inhomogeneous film. In this scenario, increasing the
magnetic field on the low-field side of the MR peak, increases the
resistance by localizing CPFs bosons in real-space mesoscopic puddles,
whereas their quantum tunneling that is followed by pair-breaking leaves the
residual inter-puddle transport to be dominated by fermionic quasi-particle
(FQP) excitations whose conductance in the spatially inhomogeneous film is
thermally activated \cite{AdkinsJPC1980}, \cite{GantmakherJETPLett95}. On
the other hand, on the high-field side of the MR peak the shrinking
real-space puddles of CPFs and the consequently expanding normal-state
regions lead to suppression of the quai-particle insulating gap and so to
decreasing magneto-resistance.

Within our quantitative analysis of the experimental sheet resistance data
reported in Ref. \cite{GantmakherJETPLett2000} for amorphous Indium-Oxide
thin films we identify the phenomenological parameter of CPF
tunneling-pair-breaking "temperature" $T_{Q}$ as representing quantum
fluctuation effects peculiar to the field-induced SIT under study. In
particular, for temperatures $T$ well below $T_{Q}$, the field-dependent
isotherms of the calculated sheet resistance were shown to have a single
crossing point at a fixed quantum critical field $H_{c}$ not far below the
MR peak, in good agreement with the experimentally observed data.

The paper is organized as follows: In Sec.II we present the theoretical
model of the disordered 2D electron system with SO scatterings subject to
pairing interaction in a parallel magnetic field. In Sec.III we develop the
theory of CPFs in the framework of this model by employing the TDGL-Lagevine
approach, which is responsible for superconductivity in the low-field region
and leads to the appearance of localized CPF bosons at high fields. In
Sec.IV we introduce the phenomenological model of field-enhanced  thermally
activated FQP transport, which leads to negative MR at high fields. The
joint effect of quantum tunneling and pair breaking of CPF bosons, which
provides the connection between the localized boson model and the FQP
transport model, is introduced phenomenologically in Sec.V. In Sec.VI we
quantitatively analyze the experimentally measured sheet resistance data
reported for amorphous Indium-Oxide thin films, by applying the hybrid
microscopic-phenomenological model developed in the preceding sections, and
finally the results are discussed and concluded in Sec.VII.

\bigskip

\section{The theoretical microscopic model}

The model employed here, which is an extension of the model presented in
Refs. \cite{MZPRB2021},\cite{MZJPC2023}, consists of a thin rectangular film
of disordered electron system, under a strong magnetic field $H$, applied
parallel to the conducting plane (see the comments concerning the situation
of perpendicular field orientation in Secs.VI and VII). Disorder is due to
impurity scatterings, through both orbital and spin-orbit scattering
potentials $V_{OR}$,$V_{SO}$, respectively, described in real-space by the
scattering matrix \cite{Abrikosov62}, \cite{Klemm75}, \cite{Fukuyama81}:%
\textbf{\ }

\begin{equation}
V\left( \mathbf{r,r}^{\prime }\right) =\frac{1}{d}\sum \limits_{n}\int \frac{%
d^{2}p}{\left( 2\pi \right) ^{2}}\int \frac{d^{2}q}{\left( 2\pi \right) ^{2}}%
\exp \left \{ i\mathbf{p\cdot }\left[ \frac{1}{2}\left( \mathbf{r}+\mathbf{r}%
^{\prime }\right) -\mathbf{R}_{n}\right] +i\mathbf{q\cdot }\left( \mathbf{r}-%
\mathbf{r}^{\prime }\right) \right \} \left( V_{OR}+iV_{SO}\left[ \mathbf{p}%
\times \mathbf{q}\right] \mathbf{\cdot \sigma }\right)  \label{V(r,r')}
\end{equation}%
where $\mathbf{\sigma}$ is the vector of the Pauli matrices,$\left(\sigma_x,\sigma_y,\sigma_z \right)$,
$\mathbf{R}_{n}$ is a position vector of an impurity and $\mathbf{p,q}$
are inplane electron wave vectors, so that $\left[ \mathbf{p}\times \mathbf{q%
}\right] \mathbf{\cdot \sigma }\neq 0$ only for electron spins perpendicular
to the conducting plane.

In the dirty limit (see below) the corresponding impurity scattering
renormalized pairing vertex factors in the imaginary (Matsubara)
frequency-wave number representation take the form \cite{MZPRB2021}:%
\begin{equation}
\widetilde{s}_{\pm }\left( \omega _{n};q,\Omega _{\nu }\right) \approx \pi
\theta \left( \omega _{n+\nu }\omega _{n}\right) \frac{\left \vert \omega
_{n}+\Omega _{\nu }/2\right \vert +\frac{1}{\tau _{SO}}+\frac{1}{2}Dq^{2}+%
\frac{1}{2}\Omega _{\nu }\pm i\left( \frac{\mu _{B}H}{\hbar }\right)
sgn\left( \omega _{n}+\Omega _{\nu }/2\right) }{\left( \left \vert \omega
_{n}+\Omega _{\nu }/2\right \vert +\frac{1}{2\tau _{SO}}+\frac{1}{2}%
Dq^{2}\right) ^{2}+\left( \frac{\mu _{B}H}{\hbar }\right) ^{2}-\left( \frac{1%
}{2\tau _{SO}}\right) ^{2}}  \label{StildePMexpl}
\end{equation}%
where $\theta(x)$ is the Heaviside step function, $\omega _{n}\equiv \left( 2n+1\right) \pi k_{B}T/\hbar ,n=0,\pm
1,\pm 2,...,\Omega _{\nu }\equiv 2\nu \pi k_{B}T/\hbar $, \  \  \ $\nu
=0,1,2,....$, $\ D\equiv \tau v_{F}^{2}/2$ is the diffusion coefficient, $%
1/\tau =1/\tau _{OR}+1/\tau _{SO}$ is the total scattering relaxation rate,
with: $\tau _{OR},\tau _{SO}$ the orbital and SO relaxation times
respectively. The dirty limit amounts to assuming that: $\hbar /\tau >>\mu
_{B}H,k_{B}T$.

Under the action of the SO component of Eq.\ref{V(r,r')}, similar to the
situation in Ising superconductors \cite{IsingSC},\cite{HaoranAccMatRes2024}%
, the Zeeman splitting pair-breaking energy vanishes to first order in $2\mu
_{B}H/\varepsilon _{SO}$, where $\varepsilon _{SO}=\hbar /\tau _{SO}$ is the
SO interaction energy. However, to second and higher orders, the
SOS-suppressed Zeeman-splitting pair-breaking effect is not negligible and
can have peculiar physical outcomes. To gain insights into the salient
features of this effect, relevant to the main subject of the paper, we
consider the fluctuations propagator, $\mathcal{D}\left( q,i\Omega _{\nu
}\right) $, after analytic continuation to real boson frequency $i\Omega
_{\nu }\rightarrow \omega $:

\begin{eqnarray}
\left[ \mathcal{D}\left( q,\omega \right) N_{2D}\right] ^{-1} &=&\varepsilon
_{H}+a_{+}\left[ \psi \left( \frac{1}{2}+f_{-}+\frac{\hbar \left(
Dq^{2}-i\omega \right) }{4\pi k_{B}T}\right) -\psi \left( \frac{1}{2}%
+f_{-}\right) \right]  \label{FlucProp} \\
&&+a_{-}\left[ \psi \left( \frac{1}{2}+f_{+}+\frac{\hbar \left(
Dq^{2}-i\omega \right) }{4\pi k_{B}T}\right) -\psi \left( \frac{1}{2}%
+f_{+}\right) \right]  \notag
\end{eqnarray}%
where:%
\begin{equation}
\varepsilon _{H}\equiv \ln \left( \frac{T}{T_{c0}}\right) +a_{+}\psi \left( 
\frac{1}{2}+f_{-}\right) +a_{-}\psi \left( \frac{1}{2}+f_{+}\right) -\psi
\left( \frac{1}{2}\right)  \label{eps_H}
\end{equation}%
and: 
\begin{equation}
a_{\pm }\equiv \frac{1}{2}\left \{ 1\pm \left[ 1-\left( \frac{2\mu _{B}H}{%
\varepsilon _{SO}}\right) ^{2}\right] ^{-1/2}\right \} ,f_{\pm }\equiv \frac{%
\varepsilon _{SO}}{4\pi k_{B}T}\left \{ 1\pm \left[ 1-\left( \frac{2\mu _{B}H%
}{\varepsilon _{SO}}\right) ^{2}\right] ^{1/2}\right \}  \label{af_pm}
\end{equation}%
(see also a detailed derivation in early papers dealing with similar model
systems \cite{Maki66}, \cite{FuldeMaki70}, \cite{WHH66}).

In these expressions $T_{c0}$ is the zero-field, mean-field transition
temperature, $\psi $ is the digamma function, and $N_{2D}=m^{\ast }/2\pi
\hbar ^{2}$ is the single electron DOS per unit area. Eqs.\ref{eps_H}, \ref%
{af_pm} show that, to first order in the the ratio $2\mu _{B}H/\varepsilon
_{SO}$, the SOS cancels the Zeeman spin-splitting pair-breaking effect,
leaving only higher order terms, quadratic in this ratio, to influence the
mean-field critical shift parameter $\varepsilon _{H}$ continuously through
the crossing point $2\mu _{B}H=\varepsilon _{SO}$.

A peculiar feature of this SOS-suppressed Zeeman splitting pair breaking
concerns its influence on the stiffness of the fluctuation modes. This can
be most clearly shown by using the small wavenumber expansion of the static
fluctuations propagator, Eq.\ref{FlucProp}:

\begin{equation}
\left[ \mathcal{D}\left( q,0\right) N_{2D}\right] ^{-1}\simeq \varepsilon
_{H}+\eta \left( H\right) \frac{\hbar D}{4\pi k_{B}T}q^{2}=\varepsilon
_{H}+\xi ^{2}\left( H\right) q^{2}  \label{smallq}
\end{equation}%
where 
\begin{equation}
\xi ^{2}\left( H\right) \equiv \eta \left( H\right) \frac{\hbar D}{4\pi
k_{B}T}  \label{xi^2(H)}
\end{equation}%
and:

\begin{equation}
\eta \left( H\right) =a_{+}\psi ^{\prime }\left( \frac{1}{2}+f_{-}\right)
+a_{-}\psi ^{\prime }\left( \frac{1}{2}+f_{+}\right)  \label{eta}
\end{equation}

It is remarkable that regardless of the magnitude of $2\mu _{B}H/\varepsilon
_{SO}$, in the low temperature limit, where $\ f_{-}\gg 1$, the reduced
stiffness parameter in Eq.\ref{eta} takes the simple limiting form (see
Appendix A):

\begin{equation}
\eta \left( H\right) \rightarrow \frac{2T}{T_{H}}  \label{etasuppr}
\end{equation}%
with the characteristic field-dependent temperature:%
\begin{equation}
T_{H}\equiv \frac{\left( \mu _{B}H\right) ^{2}}{\pi k_{B}\varepsilon _{SO}}
\label{T_H}
\end{equation}

Thus, by lowering the temperature well below $T_{H}$ the reduced stiffness
can be suppressed well below its zero field value $\eta \left( 0\right) =\pi
^{2}/2$. The condition on field and temperature for this to happen (i.e. for 
$\eta \left( H\right) /\eta \left( 0\right) =2\eta \left( H\right) /\pi
^{2}\ll 1$), is therefore:%
\begin{equation}
k_{B}T\ll \frac{1}{2\pi }\left( \frac{\mu _{B}H}{\varepsilon _{SO}}\right)
\mu _{B}H  \label{suppreta}
\end{equation}%
in addition to the much weaker condition $4\pi k_{B}T\ll \varepsilon _{SO}$.
It should be therefore noted that the condition \ref{suppreta} is valid for
a broad range of SOS energies, not necessarily restricted to the large SOS
values discussed in Refs.\cite{MZPRB2021}, \cite{MZJPC2023}, since for all
experimentally relevant field values the usual SOS energies easily satisfy $%
\varepsilon _{SO}>\mu _{B}H$.

\bigskip

\section{Field-induced localization of CPF bosons}

\bigskip

In what follows we will show how a central notion in the SIT scenario
promoted in Ref.\cite{GantmakherPhysicaB2000} can be realized within our
TDGL-Langevin approach, allowing a quantitative treatment of the
corresponding electrical transport problem. In terms of our model system
represented by the fluctuation propagator, Eq.\ref{FlucProp}, the
correlation function of the TDGL wavefunctions in momentum space (see
Appendix B) takes the form:

\begin{equation}
\left \langle \phi ^{\ast }\left( \mathbf{q};t\right) \phi \left( \mathbf{q}%
;0\right) \right \rangle =\left \langle \left \vert \phi \left( \mathbf{q}%
\right) \right \vert ^{2}\right \rangle \exp \left( -\frac{t}{\tau
_{GL}\left( q;H\right) }\right)  \label{FluctDiss}
\end{equation}%
with the momentum distribution function:

\begin{equation}
\left \langle \left \vert \phi \left( \mathbf{q}\right) \right \vert
^{2}\right \rangle =\left( \frac{7\zeta \left( 3\right) E_{F}}{4\pi
^{2}k_{B}T}\right) \frac{1}{\varepsilon _{H}+\xi ^{2}\left( H\right) q^{2}}
\label{momdistrHex}
\end{equation}%
and the corresponding life-time:%
\begin{equation}
\tau _{GL}\left( q;H\right) =\left( \frac{\eta \left( H\right) \hbar }{4\pi
k_{B}T}\right) \frac{1}{\varepsilon _{H}+\xi ^{2}\left( H\right) q^{2}}
\label{tau_GL(q;H)}
\end{equation}

Within this time scale one may define the total CPFs density per unite
volume as:%
\begin{eqnarray}
&&n_{CPF}\left( H\right) \equiv \frac{1}{d}\int \frac{d^{2}q}{\left( 2\pi
\right) ^{2}}\left \langle \left \vert \phi \left( \mathbf{q}\right) \right
\vert ^{2}\right \rangle  \notag \\
&=&\left( \frac{7\zeta \left( 3\right) E_{F}}{4\pi ^{2}k_{B}T}\right) \frac{%
\pi }{d}\int_{0}^{q_{c}^{2}}\frac{d\left( q^{2}\right) }{\left( 2\pi \right)
^{2}}\frac{1}{\varepsilon _{H}+\eta \left( H\right) \frac{\hbar D}{4\pi
k_{B}T}q^{2}}  \label{n_S(H)}
\end{eqnarray}%
where $q_{c}$ is a cutoff wavenumber.

The validity of Eq.\ref{n_S(H)} as a well-defined particle density is
limited by the life-time $\tau _{GL}\left( q;H\right) $, as given by Eq.\ref%
{tau_GL(q;H)}. At low temperatures, $T\ll T_{H}$, where:

\begin{equation*}
\tau _{GL}\left( q;H\right) \rightarrow \frac{\hbar /2\pi k_{B}T_{H}}{%
\varepsilon _{H}+\hbar Dq^{2}/2\pi k_{B}T_{H}}
\end{equation*}%
our estimate of this characteristic time at $H=8T$ (for which $T_{H}\approx
150mK$) and for the typical experimental parameters (see below) is: $\tau
_{GL}\left( q;H=8T\right) \geq \tau _{GL}\left( q_{c};H=8T\right) \approx
4\times 10^{-11}\boldsymbol{\mathit{s}}$, that is at least three orders of
magnitude longer than the typical electron relaxation time ( $\tau \approx
4\times 10^{-14}\boldsymbol{\mathit{s}}$).

These long-lived boson excitations have, in the low temperature limit, $T\ll
T_{H}$, field-enhanced coherence length:

\begin{equation}
\xi \left( H\right) \rightarrow \frac{\hbar v_{F}}{2\mu _{B}H}\sqrt{\frac{%
\tau _{OR}/\tau _{SO}}{1+\tau _{OR}/\tau _{SO}}}\sim \frac{\hbar v_{F}}{2\mu
_{B}H}  \label{xi(H)lim}
\end{equation}%
which,according to Eq.15, determines a localization length:
$\rho_{loc} \equiv \xi(H)/\sqrt{\varepsilon_{H}}$
of bosons in real-space mesoscopic puddles upon increasing magnetic field (see Ref.\cite{MZ-arXiv2025}).

The most remarkable feature of the boson density, Eq.\ref{n_S(H)}, is its
field-induced divergence in the low temperature limit, $T\ll T_{H}$, where
the reduced stiffness $\eta \left( H\right) $ (see Eq. \ref{etasuppr})
diminishes. This immediately implies that at a certain field-dependent
temperature the CPF bosons system would consume all the normal-state
electrons available for pairing, indicating the breakdown of the GL thermal
fluctuations approach. Besides this peculiarity, which is also related to
the boson localization under increasing field mentioned above, there is the
critical divergence at $\varepsilon _{H}=0$ which reflects the opposing
tendency toward long range SC order.

A plausible correction of the Gaussian approximation inherent to the
fluctuation propagator, Eq.\ref{FlucProp}, which treats the critical
divergency and allows extension of the theory well below the critical field 
\cite{MZPRB2021}, is the self-consistent field (SCF) approximation \cite%
{UllDor90}, \cite{UllDor91} of the interaction between Gaussian
fluctuations. In terms of the boson density, Eq.\ref{n_S(H)}, the
corresponding SCF equation for the "dressed" critical-shift parameter $%
\widetilde{\varepsilon }_{H}$ reads:%
\begin{equation}
\widetilde{\varepsilon }_{H}=\varepsilon _{H}+\frac{1}{2\pi }\mathcal{F}%
\left( H\right) \widetilde{n}_{CPF}\left( H;\widetilde{\varepsilon }%
_{H}\right)  \label{SCFeq}
\end{equation}%
where $\widetilde{n}_{CPF}\left( \widetilde{\varepsilon }_{H}\right) $ is
the CPFs density normalized with respect to the total number of available
electron pairs:

\begin{equation*}
\widetilde{n}_{CPF}\left( H;\widetilde{\varepsilon }_{H}\right) \equiv \frac{%
n_{CPF}\left( H\right) }{n_{0}/2},n_{0}=\frac{1}{d}\frac{k_{F}^{2}}{2\pi }
\end{equation*}%
and%
\begin{equation}
\mathcal{F}\left( H\right) =\sum \limits_{n=0}^{\infty }\frac{\varkappa _{n}%
\left[ \varkappa _{n}^{2}-3\left( \frac{\mu _{B}H}{2\pi k_{B}T}\right) ^{2}%
\right] }{\left[ \varkappa _{n}\left( \varkappa _{n}-\frac{\varepsilon _{SO}%
}{2\pi k_{B}T}\right) +\left( \frac{\mu _{B}H}{2\pi k_{B}T}\right) ^{2}%
\right] ^{3}}  \label{calF(H)}
\end{equation}%
is the interaction parameter \cite{MZPRB2021}, with: $\varkappa _{n}=n+1/2%
\mathbf{+}\varepsilon _{SO}/2\pi k_{B}T$. Performing the momentum space
integration in Eq.\ref{n_S(H)} it is found that:

\begin{equation}
\widetilde{n}_{CPF}\left( H;\widetilde{\varepsilon }_{H}\right) \simeq \frac{%
4}{\pi }\left( \frac{\hbar /\tau }{E_{F}}\right) \frac{1}{\eta \left(
H\right) }\ln \left( 1+\frac{\zeta _{c}\left( H\right) }{\widetilde{%
\varepsilon }_{H}}\right)  \label{normDens}
\end{equation}%
with the field-dependent dimensionless cutoff: $\zeta _{c}\left( H\right)
\equiv \eta \left( H\right) \hbar Dq_{c}^{2}/4\pi k_{B}T$.

Using Eq.\ref{normDens}, together with the SCF equation \ref{SCFeq}, we can
determine, at any field $H$, the temperature $T_{sat}\left( H\right) $ where
the CPF bosons system saturates the entire normal-state electrons reservoir
available for pairing, that is where $\widetilde{n}_{CPF}\left( H;\widetilde{%
\varepsilon }_{H}\right) =1$. \ At low temperatures, the SCF equation (Eq.%
\ref{SCFeq}) yields nearly vanishing $\widetilde{\varepsilon }_{H}$ at all
fields below the nominal (mean-field) critical field $H_{c}\left( T\right) $
(corresponding to $\varepsilon _{H}=0$), where $\widetilde{\varepsilon }_{H}$
crossovers asymptotically to $\varepsilon _{H}$ (see Fig.1). Thus, in the
entire fields range $0<H<H_{c}\left( T\right) $, where $-\varepsilon
_{H}=\left \vert \varepsilon _{H}\right \vert \gg \widetilde{\varepsilon }%
_{H}$, we find the simple relation:

\begin{figure}[b]
\subfloat[The "bare" (red) and the "dressed" (blue) critical shift
parameters, $\protect \varepsilon _{H}$ and $\widetilde{\protect \varepsilon }_{H}$ respectively, as functions of field calculated at two temperatures $T=100$ mK (dashed lines), and $200$ mK (solid lines).]{\includegraphics[width=0.45\columnwidth]{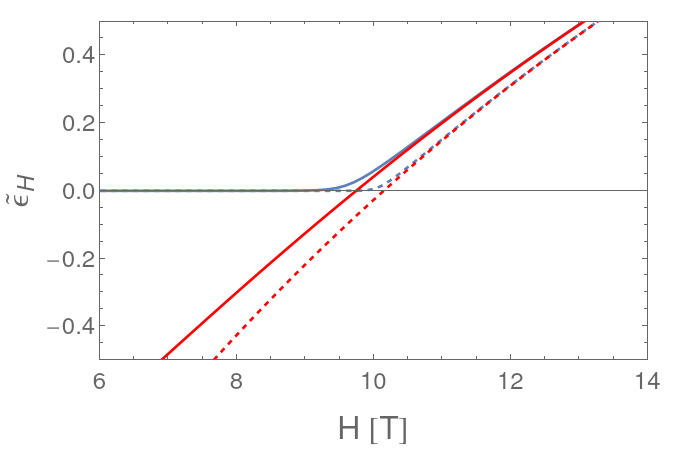}}
\hfill 
\subfloat[The mean-field
phase boundary $T_{c}\left( H\right) $, the saturation line $T_{sat}\left(
H\right) $, and the field-suppressed stiffness line $T_{H}$. ]{\includegraphics[width=0.45\columnwidth]{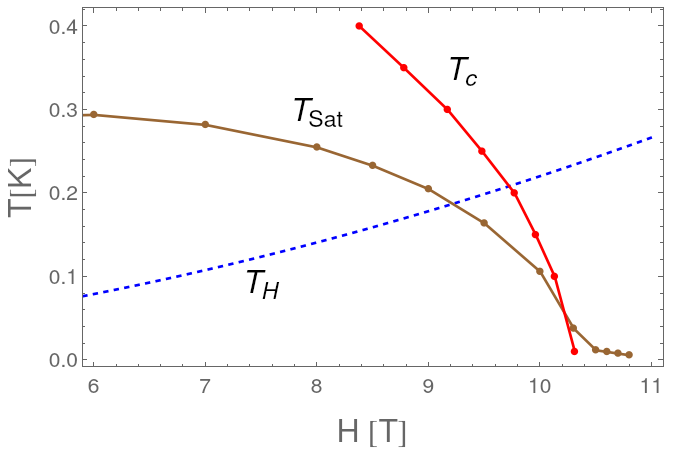}}
\caption{{}Characteristic mean-field and thermal fluctuations aspects of the
2D electron system under study. Note the low temperatures mean-field
critical field $H_{c}\left( 0\right) \sim 10$ T, in comparison with Fig.3,
where the effect of the quantum fluctuations discussed in Sec.V strongly
suppresses it toward $\sim 6$ T. For both (a) and (b) $\protect \varepsilon %
_{SO}=5.6$ meV, and: $T_{c0}=0.8$ K. \ The additional parameters used in the
calculation of $T_{sat}(H)$ are: $E_{F}=50$ meV, $\hbar /\protect \tau %
_{OR}=10$ meV, and the dimensionless cutoff parameter: $x_{c0}\equiv \hbar
Dq_{c}^{2}/4\protect \pi k_{B}T_{c0}=0.01$}
\label{fig:fig1}
\end{figure}

\begin{equation}
\widetilde{n}_{CPF}\left( H;\widetilde{\varepsilon }_{H}\right) \simeq \frac{%
2\pi \left \vert \varepsilon _{H}\right \vert }{\mathcal{F}\left( H\right) }
\label{normDenSimp}
\end{equation}%
so that the saturation temperature $T_{sat}\left( H\right) $ can be
approximately determined from the equation: $2\pi \left \vert \varepsilon
_{H}\right \vert =\mathcal{F}\left( H\right) $.

In Fig.1 we plot both the saturation line, $T_{sat}\left( H\right) $, and
the mean-field phase boundary, $T_{c}\left( H\right) $, together with $T_{H}$
(Eq.\ref{T_H})-the characteristic temperature below which the reduced
stiffness is significantly suppressed with respect to its zero field value $%
\eta \left( 0\right) =\pi ^{2}/2$. It should be emphasized that the
saturation line, $T_{sat}\left( H\right) $, is determined mainly by the
interplay between the "bare" critical shift and the interaction parameters, $%
\varepsilon _{H}$ and $\mathcal{F}\left( H\right) $, respectively, and is
independent of both the Fermi energy $E_{F}$ and the reduced stiffness
parameter $\eta \left( H\right) $, appearing in Eq.\ref{normDens}, except
for the close vicinity of $H_{c}\left( T\rightarrow 0\right) $, where $%
\widetilde{\varepsilon }_{H}\approx \left \vert \varepsilon _{H}\right \vert 
$ and $T_{sat}\left( H\right) \approx T_{c}\left( H\right) $ (see Fig.1).

It is also interesting to note that these two lines approach each other not
too far above $H_{c}\left( T\rightarrow 0\right) $, at temperatures well
below $T_{H}$, where the suppressed reduced stiffness parameter $\eta \left(
H\right) $ leads to localized condensation of CPF bosons in real space.

The corresponding (para) conductivity of the bosons system, which is
equivalent to the Aslamazov-Larkin conductivity \cite{AL68} in the
microscopic theory, is written in the form \cite{MZ-arXiv2025}:

\begin{equation}
\sigma _{AL}\simeq \left( \frac{e^{2}}{16\hbar d}\right) \frac{\eta \left(
H\right) }{\eta \left( 0\right) }\frac{1}{\widetilde{\varepsilon }_{H}\left(
1+\frac{\widetilde{\varepsilon }_{H}}{\zeta _{c}\left( H\right) }\right) }
\label{sig_AL}
\end{equation}

This expression reflects the two competing effects, discussed above with
regards to the CPFs density Eq.\ref{n_S(H)}, on the boson transport : On the
one hand the long-range SC order that is developed as $\widetilde{%
\varepsilon }_{H}$ tends to zero under diminishing field, and on the other
hand the very sluggish CPFs transport which takes place due to the sharp
suppression of the reduced stiffness $\eta \left( H\right) $ (see Eq. \ref%
{etasuppr}) at high fields.

\section{Field-enhanced thermally activated fermionic quasi-particle
transport}

In the microscopic (BCS) theory of superconductivity the underlying
electrons system is usually treated as a grand canonical ensemble, which is
a justified assumption only when the number of excited CPFs is much smaller
than half of the total number of electrons available for pairing. As
discussed in detail in the preceding section, this condition is evidently
not satisfied in our model system at low temperatures and finite magnetic
field; $T\leq T_{sat}\left( H\right) $, where $\widetilde{n}_{CPF}\left( 
\widetilde{\varepsilon }_{H}\right) \geq 1$, so that one expects the
microscopic transport theory of SC fluctuations \cite{LV05} to lose validity
under these circumstances.

Furthermore, for the low carrier densities characterizing the model systems
under study the prefactors $\hbar /\tau E_{F}$ in Eq.\ref{normDens} is much
larger than for those of good metals so that the breakdown of the
conventional microscopic theory is expected to occur at experimentally
attainable temperatures. Besides we find that, consistently with the CPF
bosons localization in real-space mesoscopic puddles predicted in our
theory, transport measurements on the disordered 2D electron systems under
study and their analyses reported in Refs. \cite{DynesPRL1978},\cite%
{AdkinsJPC1980} have indicated that they form spatially inhomogeneous
structures consisting of SC grains embedded in highly resistive media \cite%
{GerberPRL1997},\cite{BeloborodovPRL1999},\cite{BeloborodovPRB2000}. The
observation of field-enhanced thermally activated transport in, e.g. thin
amorphous Indium Oxide films \cite{GantmakherJETP96} has been associated
with inter-grain tunneling of FQPs rather than with Josephson tunneling \cite%
{AdkinsJPC1980}, similar to the situation in granular 3D superconductors 
\cite{GerberPRL1997},\cite{BeloborodovPRB2000}.

This observation is fully consistent with our phenomenological introduction
of joint effect of quantum tunneling of CPFs and their pair breaking that
prevent over-saturation of CPFs puddles (see the next section). In
particular, the pair-breaking process that consecutively follows the
tunneling of CPFs bosons out of their mesoscopic enclaves reinforces
inter-puddle transport of FQPs over Josephson tunneling of Cooper pairs.
Thus, one expects the FQP activation gap to be suppressed at high fields
since under increasing magnetic field the generating inhomogeneous
structures disappear with the diminishing CPFs density in shrinking
mesoscopic puddles. For example, at large field above the SC critical field,
the CPFs density is suppressed due to the overwhelming effect of the
critical shift parameter $\widetilde{\varepsilon }_{H}$ in Eq. \ref{normDens}
with respect to the opposing effect of the stiffness parameter $\eta \left(
H\right) $, whereas the CPF localization length (Eq.\ref{xi^2(H)})
diminishes with $\eta \left( H\right) $.

Thus, following the ansatz proposed in Ref.\cite{GantmakherJETP96} we write
the residual FQP conductivity in the thermally activated form: 
\begin{equation}
\sigma _{FQP}\left( T,H\right) =\sigma _{n0}\left( T\right) \exp \left[ -%
\frac{\Delta \left( h\right) }{k_{B}T}\right]  \label{sig_FQP}
\end{equation}%
where the field-dependent energy-gap is given by:%
\begin{equation}
\Delta \left( h\right) =\frac{\Delta _{0}}{h}  \label{Del(h)}
\end{equation}%
$h\equiv H/H_{0}$, $H_{0}$ is a characteristic field strength, and $\Delta
_{0}$ is a characteristic energy gap at $H=H_{0}$.

Defining an effective temperature-dependent magnetic field activation gap:%
\begin{equation}
H_{Gap}\left( T\right) \equiv \frac{H_{0}\Delta _{0}}{k_{B}T}  \label{H_Gap}
\end{equation}%
Eq.\ref{sig_FQP} is rewritten in the form:

\begin{equation}
\sigma _{FQP}\left( T,H\right) =\sigma _{n0}\left( T\right) \exp \left[ -%
\frac{H_{Gap}\left( T\right) }{H}\right]  \label{sig_FQPasymp}
\end{equation}

\bigskip

\section{Quantum tunneling and pair breaking of CPF bosons}

\bigskip

As discussed in the preceding section, the oversaturation ($\widetilde{n}%
_{CPF}\left( \widetilde{\varepsilon }_{H}\right) >1$) of CPF bosons in
mesoscopic puddles at low temperatures $T$ and finite magnetic fields $H$
satisfying $T<T_{sat}\left( H\right) $, is a clear indication of the
invalidity of the grand canonical ensemble underlying the microscopic theory
of fluctuations in superconductors \cite{LV05}. A physically plausible
correction to this nonphysical deviation has been originally proposed in 
\cite{MZPRB2021}, \cite{MZJPC2023}, in which the overfilling CPFs bosons are
forced by their increasing "osmotic pressure" to tunnel out of their
mesoscopic enclaves and then instantaneously pair-break into unpaired
fermionic quasi particles.

Formally speaking, the quantum tunneling processes of CPFs lead to a
recovery of the vanishing reduced stiffness parameter $\eta \left( H\right) $
in the zero temperature limit, which can prevent the oversaturation of CPF
bosons. This is shown (see Appendix C) within a simple model hamiltonian of
bosons, whose energy dispersion in momentum space follows that of our CPFs,
which aggregate into a 2D network of localized puddles arranged in 2D
disordered lattice. The resulting tunneling-induced modification of the
quadratic energy dispersion is equivalent to a non-vanishing shift $\Delta
\eta _{tunn}$ of the reduced stiffness function $\eta \left( H\right) $ (see
Appendix C), which can be effectively described in terms of a tunneling
temperature $T_{Q}$ by the replacement formula: 
\begin{equation}
\eta \left( H\right) \rightarrow \eta \left( H\right) +\Delta \eta
_{tunn}\equiv \left( 1+T_{Q}/T\right) \eta \left( H\right)
\label{eta(H)correcTq}
\end{equation}

In this expression the singular $T_{Q}/T$ term of the correction factor
exactly cancel the vanishing stiffness in the low temperature limit, $T\ll
T_{H}$, so that $\eta \left( H\right) \rightarrow \Delta \eta
_{tunn}\rightarrow 2T_{Q}/T_{H}$.

In Appendix D it is shown that this tunneling-induced correction to $\eta
\left( H\right) $ is equivalent to implementing phenomenologically the
effect of tunneling into the statistical mechanics of the CPF bosons system
by replacing the imaginary (thermal) time interval $\tau _{T}=\hbar /k_{B}T$
in the corresponding partition function with the imaginary (quantum-thermal)
time interval $\tau _{U}=\tau _{T}\tau _{Q}/\left( \tau _{T}+\tau
_{Q}\right) =\hbar /k_{B}\left( T+T_{Q}\right) $, where $\tau _{Q}=\hbar
/k_{B}T_{Q}$ is the quantum tunneling time interval. One notes, however,
that at zero field this procedure breaks down since $\eta \left( H=0\right)
=\pi ^{2}/2$ even for $T\rightarrow 0$, so that the corrected reduced
stiffness parameter diverges at $H=0$ when $T\rightarrow 0$. Physically
speaking, one also notes that the suppression of the CPFs density via
quantum tunneling out of their mesoscopic enclaves should be consecutively
followed by appropriate introduction of the effect of quantum pair-breaking
processes, in order to conserve the total number of electrons available for
Cooper pairing.

Thus, formally, the quantum pair-breaking effect is introduced
phenomenologically into the many-electron correlation functions involved in
the calculation of the coefficients of the GL functional (see Appendix D) to
compensate for the quantum tunneling effect introduced to the collective
boson modes in terms of the tunneling time interval $\tau _{Q}=\hbar
/k_{B}T_{Q}$. The appropriate modification introduced to the two-electron
(pair) correlation function corresponds to a shift $T_{Q}/2T$ of the
argument of the digamma function in Eq.\ref{eps_H}, that is:

\begin{equation}
\varepsilon _{H}^{U}\equiv \ln \left( \frac{T}{T_{c0}}\right) +a_{+}\psi %
\left[ \frac{1}{2}\left( 1+\frac{T_{Q}}{T}\right) +f_{-}\right] +a_{-}\psi %
\left[ \frac{1}{2}\left( 1+\frac{T_{Q}}{T}\right) +f_{+}\right] -\psi \left( 
\frac{1}{2}\right)  \label{eps_H^U}
\end{equation}%
which automatically yields the corresponding modification in the reduced
stiffness parameter, Eq.\ref{eta}, namely:

\begin{equation}
\eta _{U}\left( H\right) \equiv \left( \varepsilon _{H}^{U}\right) ^{\prime
}=a_{+}\psi ^{\prime }\left[ \frac{1}{2}\left( 1+\frac{T_{Q}}{T}\right)
+f_{-}\right] +a_{-}\psi ^{\prime }\left[ \frac{1}{2}\left( 1+\frac{T_{Q}}{T}%
\right) +f_{+}\right]  \label{eta_U}
\end{equation}

Thus, the combined quantum tunneling-pair-breaking effect on the reduced
stiffness is introduced by the product: 
\begin{equation}
\Theta \left( H;T_{Q}\right) \equiv \left( 1+\frac{T_{Q}}{T}\right) \eta
_{U}\left( H\right)  \label{Theta}
\end{equation}%
(see Appendix D). In Fig.2 it is shown as a function of field $H$ at two
temperatures, $T=30,60$ mK for various values of $T_{Q}$. The influence of
increasing $T_{Q}$ on $\Theta \left( H;T_{Q}\right) $ is seen at a crossing
point $H_{cross}$ to be divided into two distinct field regimes where $%
\Theta \left( H;T_{Q}\right) $ crossovers from decreasing to increasing
function of $T_{Q}$. Except for small $T_{Q}$ values relative to $T$, the
crossing point is obtained, for all other values of $T_{Q}$, very close to $%
T_{H}=2T$, where the effective reduced stiffness is $\Theta \left(
H_{cross};T_{Q}\right) \approx 2$ (see Fig.2 and Appendix A).

\begin{figure}[b]
\includegraphics[width=0.8\columnwidth]{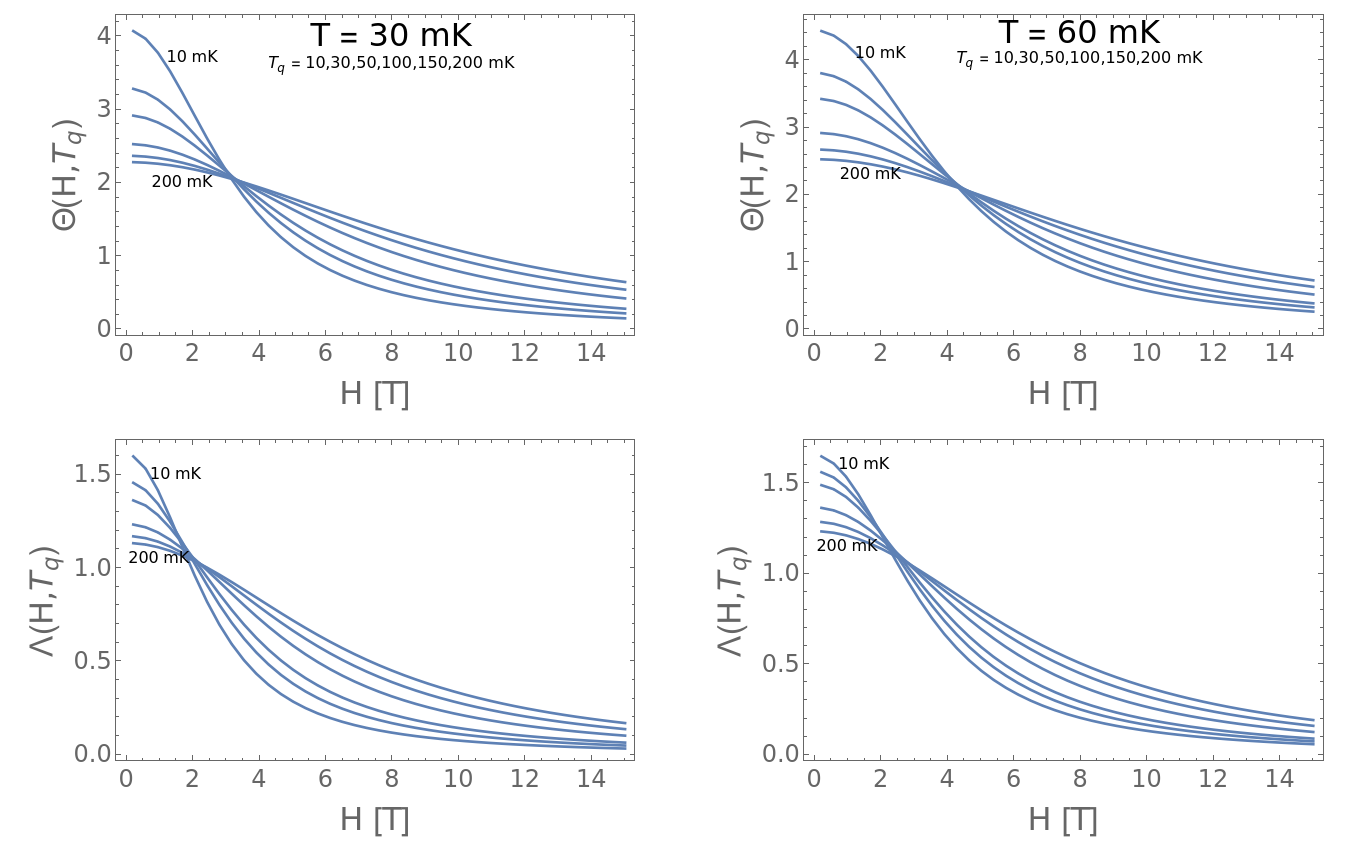}
\caption{ Upper panels: The tunneling-pair-breaking modified rduced stifness 
$\Theta \left( H;T_{Q}\right) $ (Eq.\protect \ref{Theta}) at two temperatures
calculated for various $T_{Q}$ values. Lower panels: The corresponding
tunneling-pair-breaking modified interaction parameter $\Lambda \left(
H;T_{Q}\right) $ (Eq.\protect \ref{Lambda}) at the same two temperatures and
for the same $T_{Q}$ values as in the upper panels. }
\label{fig:fig2}
\end{figure}

The increasing dependence of $\Theta \left( H;T_{Q}\right) $ on $T_{Q}$
above $H_{cross}$ reflects the fact that, at high fields, the stiffness
enhancement due to the tunneling effect overcomes the suppression associated
with the pair breaking effect, whereas its decreasing $T_{Q}$ dependence
below $H_{cross}$ reflects just the opposite counteraction situation at low
fields.

Introducing the joint tunneling-pair-breaking effect into the coupled
equations \ref{normDens} and \ref{SCFeq}, the modified equation for $%
\widetilde{n}_{CPF}^{U}$ and the SCF equation for $\widetilde{\varepsilon }%
_{H}^{U}$ take the respective forms (see Appendix D):

\begin{equation}
\widetilde{n}_{CPF}^{U}\left( H;\widetilde{\varepsilon }_{H}\right) \simeq 
\frac{4}{\pi }\left( \frac{\hbar /\tau }{E_{F}}\right) \frac{1}{\left( 1+%
\frac{T_{Q}}{T}\right) \eta _{U}\left( H\right) }\ln \left( 1+\frac{\zeta
_{c}\left( H\right) }{\widetilde{\varepsilon }_{H}^{U}}\right)
\label{normDens^U}
\end{equation}

\begin{equation}
\widetilde{\varepsilon }_{H}^{U}=\varepsilon _{H}^{U}+\left( \frac{1}{2\pi }%
\right) \left( 1+\frac{T_{Q}}{T}\right) ^{2}\mathcal{F}_{U}\left( H\right) 
\widetilde{n}_{CPF}^{U}\left( H;\widetilde{\varepsilon }_{H}^{U}\right)
\label{SCF_Ueq}
\end{equation}%
where $\mathcal{F}_{U}\left( H\right) $ is given in Appendix D (Eq.\ref%
{F_U(H)} there). In the alternative form of the SCF equation derived in
Appendix D (Eq.\ref{SCFeq^U}) one may identify the products function:%
\begin{equation}
\Lambda \left( H;T_{Q}\right) \equiv \left( 1+\frac{T_{Q}}{T}\right) \frac{%
\mathcal{F}_{U}\left( H\right) }{\eta _{U}\left( H\right) }  \label{Lambda}
\end{equation}%
in the interaction term as a higher-order analogue of $\Theta \left(
H;T_{Q}\right) $ in revealing the balance between the tunneling and the
pair-breaking effects (see Fig.2).

Finally, the modified expression for the CPF bosons paraconductivity
(compare Eq.\ref{sig_AL}) is given by:

\begin{equation}
\sigma _{AL}^{U}\left( T,H\right) \simeq \left( \frac{e^{2}}{16\hbar d}%
\right) \frac{\left( 1+\frac{T_{Q}}{T}\right) \eta _{U}\left( H\right) }{%
\eta \left( 0\right) }\frac{1}{\widetilde{\varepsilon }_{H}^{U}\left( 1+%
\frac{\widetilde{\varepsilon }_{H}^{U}}{\zeta _{c}\left( H\right) }\right) }
\label{sig_AL^U}
\end{equation}

Note that the dimensionless cutoff parameter $\zeta _{c}\left( H\right) $,
appearing in Eqs.\ref{normDens^U}, \ref{sig_AL^U}, is not affected by the
tunneling-pair-breaking procedure due to the universality of the energy
denominator, Eq.\ref{smallq}, written under the corresponding integrals over 
$q^{2}$ as a function of the dimensionless variable: $\zeta \equiv \eta
\left( H\right) \hbar Dq^{2}/4\pi k_{B}T\longleftrightarrow \left( \eta
\left( H\right) +\Delta \eta _{tunn}\right) \hbar Dq^{2}/4\pi k_{B}T$ (see
Appendix E).

\bigskip

\section{Quantitative analysis of experimental data from amorphous
Indium-Oxide thin film}

\bigskip

In this section we will use the theory presented in the preceding sections
to quantitatively analyze experimental sheet resistance data from amorphous
Indium Oxide thin films reported in Ref.\cite{GantmakherJETPLett2000}. We
focus on field orientation parallel to the film broad face in order to show
that the boson-vortex duality \cite{Fisher90}, used commonly in the field,
is not a necessary condition for the observed field-induced SIT.

\begin{figure}[tbh]
\subfloat[]{\includegraphics[width=0.48\columnwidth]{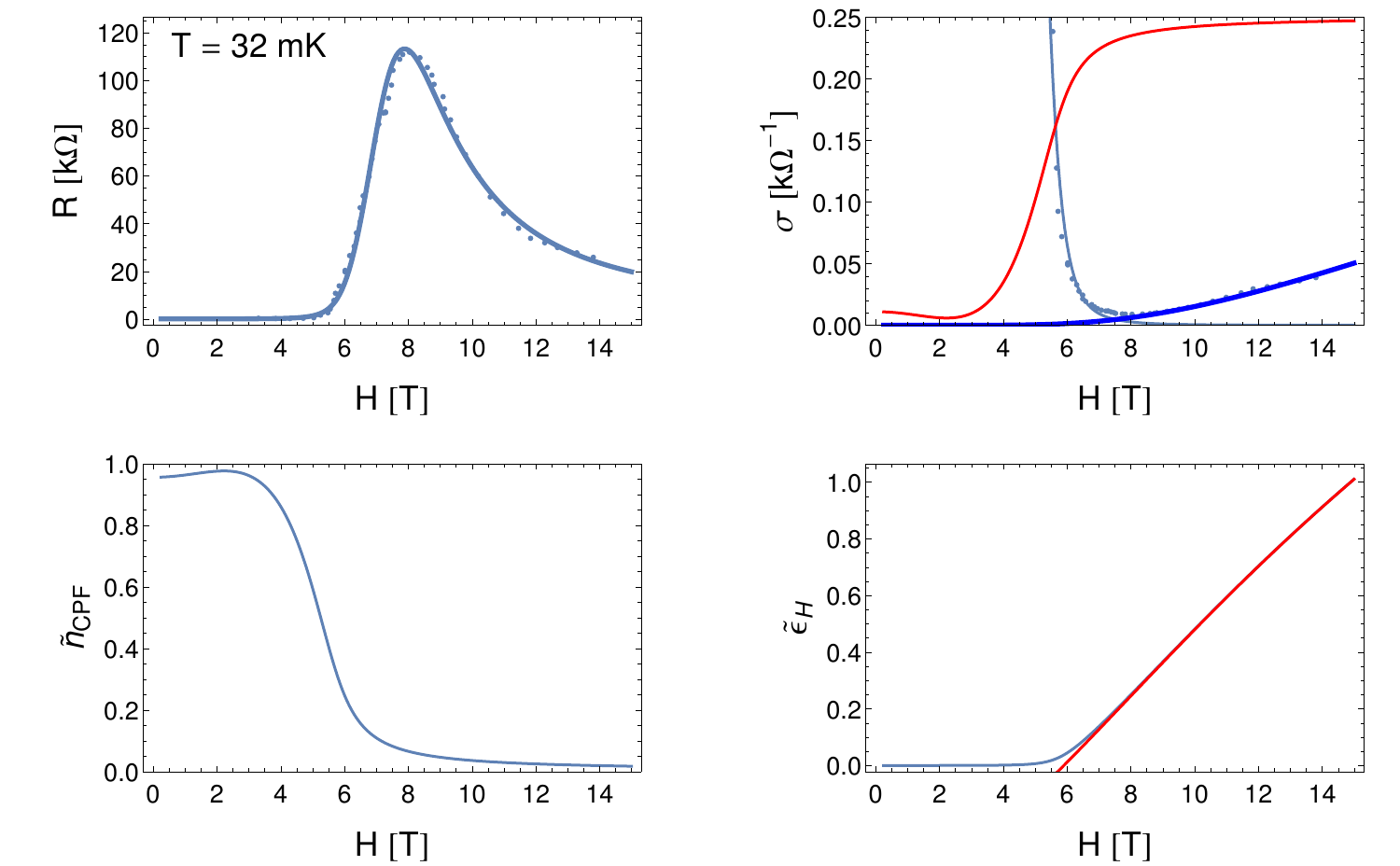}} \hfill %
\subfloat[]{\includegraphics[width=0.48\columnwidth]{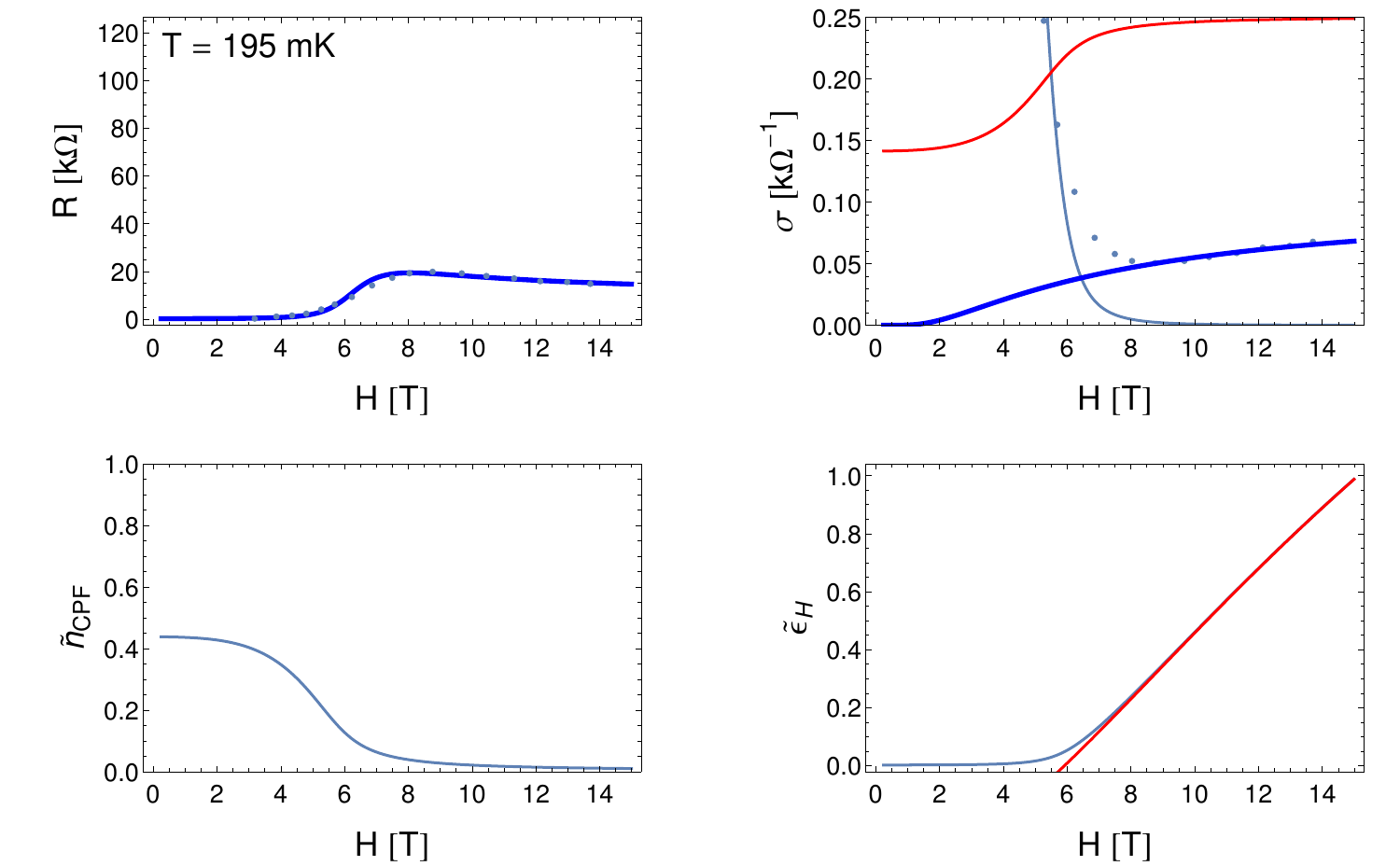}}
\caption{ Illustration of the first stage of the fitting process described
in the text at two representative temperatures; $T=32$ mK (a) and $T=195$ mK
(b). Upper-left panel in both (a) and (b): The calculated field-dependent
sheet resistance, plotted (solid line) together with the corresponding
experimental sheet resistance data (dotes). Upper-right panel: The
calculated paraconductivity (thin blue solid line) and the FQP conductivity
(thick blue solid line), plotted together with the corresponding
experimentally measured sheet conductance data (dotes) as functions of the
field. Also shown in each upper-right panel is the corresponding effective
(normal-state) Drude conductivity $\protect \sigma _{n}^{Drude}\left(
H\right) $ as influenced by the CPFs density shown in each lower-left panel.
Lower panels: The corresponding calculated normalized CPFs density $%
\widetilde{n}_{CPF}^{U}\left( H;\widetilde{\protect \varepsilon }%
_{H}^{U}\right) $ (left panel) and the "bare" (red) and "dressed" (blue)
critical shift parameters; $\protect \varepsilon _{H}^{U}$ and $\widetilde{%
\protect \varepsilon }_{H}^{U}$, respectively (right panel). }
\label{fig:fig3}
\end{figure}

The microscopic parameters, used in this analysis, have been adopted from
the source data references, \cite{GantmakherJETPLett2000}, and \cite%
{GantmakherJETP96}, as well as from complementary references (see below and an extended discussion in Ref.\cite{ManivPRB2026}).
Using the experimentally reported film thickness ($d=20$ nm) and the 3D
carrier density $n_{3D}\sim 10^{19}$ cm$^{-3}$from Ref.\cite{YeomJMC2016},
we find for the sheet (2D) density: $n_{2D}=dn_{3D}=2\times 10^{13}$ cm$%
^{-2} $, so that for a free electron mass, the Fermi energy is: $E_{F}\sim
50 $ meV. The corresponding low-temperature ($T\ll T_{H}$) coherence length (see Eq.\ref{xi(H)lim}): $\xi \left( H\right)
\rightarrow \left( \hbar v_{F}/2\mu _{B}H\right) \left( \tau \varepsilon
_{SO}/\hbar \right) \sim v_{F}\hbar /2\mu _{B}H$, at the MR peak ($H\sim 8$
T), is about $100$ nm. This relatively large value, on the scale of the film
thickness, supports our assumption of using a 2D model for the CPF bosons.
The corresponding relatively large cyclotron radius at the Fermi energy: $%
r_{F}\simeq 70$ nm is consistent with the assumption of 2D single electron
dynamics, as well. The SOS time $\tau _{SO}$ will be considered in our
analysis as an adjustable parameter, restricted however within the expected
range of $\left( 10^{-12}-10^{-13}\right) $ \textit{s} \cite{ShapirPRB1989}
that is equivalent to $\varepsilon _{SO}\sim \left( 1-10\right) $ meV. For
the zero-field SC transition temperature we use: $T_{c0}=0.8$ K \cite{GantmakherJETP96}

The total sheet magnetoconductivity of our model system of bosonic CPFs and
FQPs is now written in the form:

\begin{equation}
\sigma _{sheet}\left( T,H\right) =\sigma _{AL}^{U}\left( T,H\right) +\sigma
_{FQP}\left( T,H\right)  \label{sig(T,H)}
\end{equation}%
where $\sigma _{AL}^{U}\left( T,H\right) $ is given by Eq.\ref{sig_AL^U} and 
$\sigma _{FQP}\left( T,H\right) $ by Eq.\ref{sig_FQP}. One may divide the
entire magnetic fields region into three subregions: The low and the high
fields asymptotic regions and the intermediate region. In the low-fields
asymptotic region, $H<8$ T, the sheet conductivity is dominated by the boson
(AL) paraconductivity $\sigma _{AL}^{U}\left( T,H\right) $, approaching
superconductivity at zero field, whereas in the high-fields asymptotic
region, $H>8$ T, it is dominated by the FQP conductivity $\sigma
_{FQP}\left( T,H\right) $, given by Eq.\ref{sig_FQPasymp}, approaching from
below a poorly conductive state, i.e. a conductance small on the scale of
the metallic Drude conductivity (see Fig.3).

These functions are plotted in Fig.3 for two representative temperatures,
together with the corresponding experimental sheet conductance data
extracted from Ref.\cite{GantmakherJETPLett2000}. The fitting process has
been performed in two consecutive stages: In the first stage, the fitting is
done independently in the low and the high fields asymptotic regions,
yielding the best fitting parameters for $\sigma _{AL}^{U}\left( T,H\right) $
in the low-fields region: $\varepsilon _{SO}=5.4$ meV$,\hbar /\tau _{OR}=10$
meV and $T_{Q}$ in the range between $148$ mK to $125$ mK, depending on
temperatue, whereas for $\sigma _{FQP}\left( T,H\right) $ in the high-fields
region: $H_{0}\Delta _{0}/k_{B}=1.08[$T$\times $K$]$ with the values of $%
\sigma _{n0}\left( T\right) $ shown in Fig.4a. The thermally activated
behavior of $\sigma _{FQP}\left( T,H\right) $ is verified in the fitting
process by checking that $H_{Gap}\left( T\right) \propto 1/T$ (see Fig.4b).

\begin{figure}[tbh]
\subfloat[The pre-exponential factor of the FQP conductivity $\sigma_{n0}
(T)$ (blue full circles), plotted together with the Drude conductivity
$\sigma_{n}^{Drude}(H)$ at $H=2 T$(red full circles) and at $H=12 T$ (red
dashed line). ]{\includegraphics[width=0.45\columnwidth]{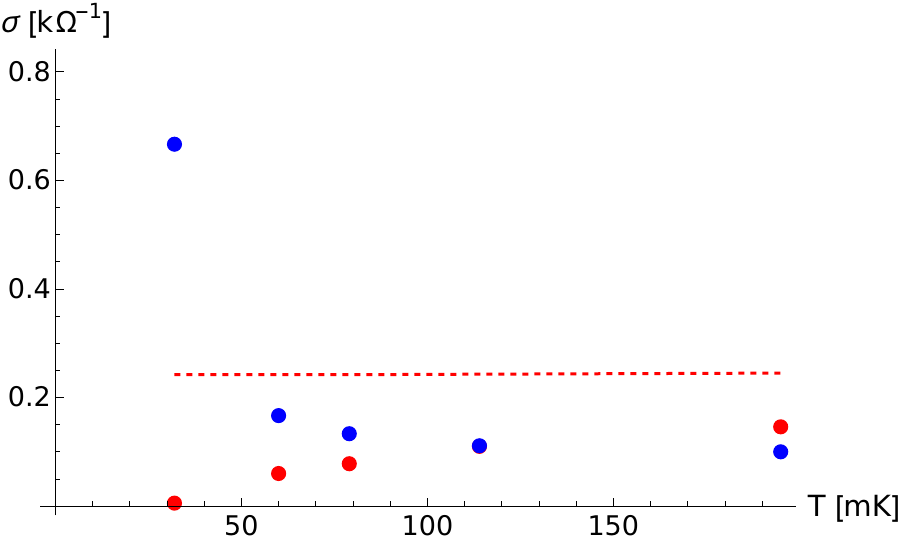}} \hfill 
\subfloat[The effective magnetic field activation gap $H_{Gap}\left( T\right) $ as a
function of temperature (blue full circles) obtained by fitting Eq.\protect
\ref{sig_FQPasymp} to the high-field experimentally measured sheet
conductance data. The red dashed line represents the fitting function $1.08/T$.]{\includegraphics[width=0.45\columnwidth]{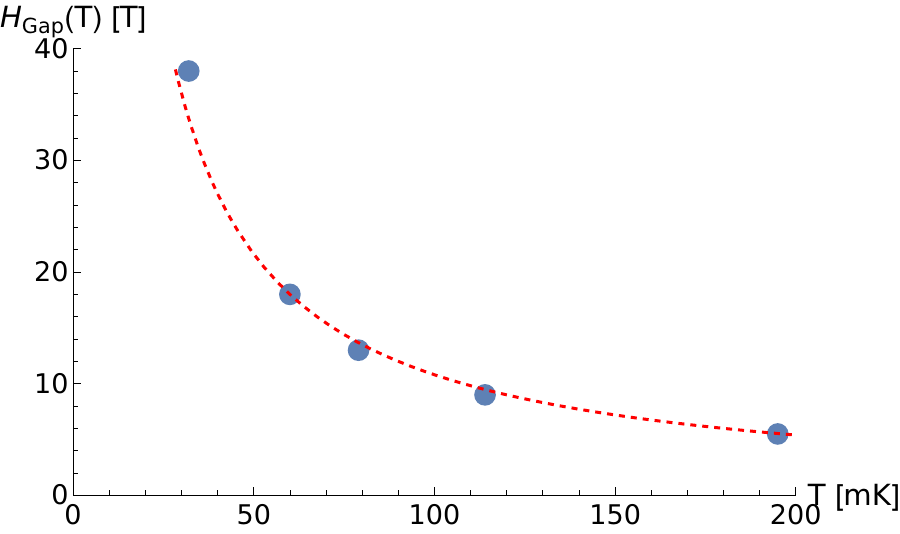}}
\caption{The FQP conductivity ingredients used in the first stage of the
fitting process, plotted as fuctions of temperature. }
\label{fig:fig4}
\end{figure}

The normalized CPFs density $\widetilde{n}_{CPF}^{U}\left( H;\widetilde{%
\varepsilon }_{H}^{U}\right) $ (Eq.\ref{normDens^U}) is used in this
analysis to oversee the selections of the relevant adjustable parameters,
particularly the tunneling-pair-breaking rate constant $\propto T_{Q}$, so
that it would not reach the over-saturation condition ($\widetilde{n}%
_{CPF}^{U}\left( H;\widetilde{\varepsilon }_{H}^{U}\right) >1$). Thus,
indeed, without actually imposing this condition, our final fitting
procedure has never failed in this important test (see Fig.3). The resulting
function $\widetilde{n}_{CPF}^{U}\left( H;\widetilde{\varepsilon }%
_{H}\right) $ is also used for calculating another reference function, the
Drude normal-state conductivity:

\begin{equation}
\sigma _{n}^{Drude}\left( H\right) =\frac{\tau e^{2}}{m^{\ast }} \left[ 1-\widetilde{n}_{CPF}^{U}\left( H;\widetilde{\varepsilon }%
_{H}^{U}\right) \right] n_{0}  \label{sig_n^Drude}
\end{equation}%
which is found in Fig.3 to be significantly larger than the FQP conductivity 
$\sigma _{FQP}\left( T,H\right) $ in the entire fields range for all
temperatures. In particular, even at very low temperatures, where the
pre-exponential factor, $\sigma _{n0}\left( T\right) $ is larger than $%
\sigma _{n}^{Drude}\left( H\right) $ (see Fig.4a), the overall activated FQP
conductivity $\sigma _{FQP}\left( T,H\right) $ remains well below $\sigma
_{n}^{Drude}\left( H\right) $. This finding reflects the poorly metallic
nature of the underlying 2D electron system in the amorphous InO film under
study.

In Fig.3 we then plot the resulting calculated total sheet resistance $%
1/\sigma _{sheet}\left( T,H\right) $ for two temperatures: $T=32$ and $195$
mK, representing the results of our calculations at various temperatures
between these two values, together with the corresponding experimentally
measured sheet resistance data reported in \cite{GantmakherJETPLett2000}.
The good agreement between theory and experiment seen in the intermediate
SIT region was obtained without any further variations of the adjustable
parameters which independently determined the asymptotic behaviors of $%
\sigma _{AL}^{U}\left( T,H\right) $ and $\sigma _{FQP}\left( T,H\right) $.

In the second stage of the fitting process salient features of the sheet
resistance data in the intermediate region are imposed on the calculated
resistance for fine tuning of the adjustable parameters. In this regard it
is interesting to note that, despite the very significant change of the
sheet resistance profile shown in Fig.3 upon decreasing temperature, the
critical field $H_{c}^{U}\left( T\right) $ (determined by $\varepsilon
_{H}^{U}=0$) is seen to be nearly independent of temperature; $H_{c}^{U}\sim
6$T. The reason for this is the relatively large quantum
tunneling-pair-breaking rate constant $\propto T_{Q}\sim 140$ mK on the
scale of the relevant temperatures $T$. In the SCF approximation used for
calculating $\widetilde{\varepsilon }_{H}^{U}$, it vanishes only at $H=0$,
remains nearly zero at finite, low fields and crossovers around $%
H_{c}^{U}\sim 6$ T asymptotically to $\varepsilon _{H}^{U}$ at high fields
(see Fig.3 and compare to Fig.1 where $H_{c}\left( T\rightarrow 0\right)
\sim 10$ T). In the crossover region the $1/\widetilde{\varepsilon }_{H}^{U}$
factor controls the field dependence of $\sigma _{sheet}\left( T,H\right) $
through the paraconductivity, Eq.\ref{sig_AL^U}, so that all the isotherms $%
\sigma _{sheet}\left( T,H\right) $ are expected to approach each other
around $H_{c}^{U}\sim 6$T provided $T\ll T_{Q}\sim 148$ mK.

Under this condition quantum criticality can emerge in our theory at low
temperatures as a natural outcome of the quantum-tunneling-pair-breaking
postulate where deviations from a single crossing of the sheet resistance
isotherms are expected upon increasing temperature. Thus, using the
experimentally observed critical (crossing) field ($H_{c}^{U}=5.5$ T) as a
fixed given parameter we repeat the fitting process by allowing the relevant
parameters $\varepsilon _{SO}$ and $T_{Q}$ to vary without changing the
other parameters. A nearly single critical (crossing) point is found at the
expected critical field (see Fig.5a) by reducing significantly the variation
of $T_{Q}$ around $T_{Q}=148$ mK in the low temperature region ($T=32,60,79$
mK), under a small modification of the SOS energy; $\varepsilon
_{SO}=5.4\rightarrow 5.6$ meV (compare Fig.5b to 5a). Note, in contrast, the
multiple crossing of the isotherms shown in Fig.5b, which was obtained in
the first stage of the fitting process summarized in Fig.3 for significantly
broader distribution of $T_{Q}$ values over temperature. The resulting
overall calculated sheet resistance curves for the different temperatures
shown in Fig.5 are seen to be in good quantitative agreement with the
experimental data reported in \cite{GantmakherJETPLett2000}.

\bigskip

\section{Discussion and conclusion}

\bigskip

The above quantitative analysis has clearly supported the proposed scenario
of the field-induced SIT observed in amorphous-Indium Oxide thin film \cite%
{GantmakherJETPLett2000}, according to which the interplay between
field-induced CPFs boson localization and field-enhanced activation of FQP
transport leads, at low temperatures, to a crossover between
superconductivity at low field and large MR peak at high field. The spirit
of this scenario is similar to that proposed in a series of papers by
Gantmakher et al. \cite{GantmakherPhysicaB2000}, however, beside the seldom
quantitative nature of the present work, there are three important new
features of our approach which significantly upgrade its potential impact.

It, first of all, introduces a concrete physical model, the localized CPFs,
behind the generic concept of localized boson system proposed in Ref.\cite%
{GantmakherPhysicaB2000}, which as indicated above, allows the quantitative
analysis of the experimental data. Second, it provides supporting evidence
for the dual (boson-fermion) nature of the proposed scenario by showing that
the tendency of the CPFs bosons to oversaturation at low temperatures should
lead, through joint process of quantum tunneling-pair-breaking to the
dominance of FQPs in the inter-puddle transport at high fields.

And last, but not the least, the phenomenological introduction of the hybrid
notion of quantum tunneling-pair-breaking mechanism into the TDGL approach
reveals a concrete physical origin of the quantum criticality observed
experimentally in the magneto-transport of amorphous Indium Oxide thin films 
\cite{GantmakherPhysicaB2000}, as well as in other related materials \cite%
{HaoranAccMatRes2024}.

The important role played in the proposed scenario by SOS, which was
discussed in some detail in Sec.II, should be reemphasized here. In
suppressing the Zeeman spin splitting pair breaking effect SOS allows
spin-singlet superconducting order to appear within a significant magnetic
fields range; $0<H<H_{c}^{U}$. However, since this SOS-suppressed
spin-splitting effect is restricted up to second order terms in $2\mu
_{B}H/\varepsilon _{SO}$, the residual, high-order pair-breaking effect is
sufficiently effective to dramatically diminish the stiffness of the
fluctuation modes at low temperatures below $T_{H}$, a phenomenon behind the
formation of boson insulating states in our theory.

Our theory is applied here only to magnetic field orientations parallel to
the conducting plane. This restricted selection has two important reasons:
1) To focus on field orientation that does not support magnetic flux
quantization, and so allowing for inquiring into the possibility that the
field-induced SIT phenomenon under study is not necessarily associated with
the boson-vortex duality \cite{Fisher90}, used commonly in the field. 2) To
avoid the complexity involved in introducing extrinsic flux-depinning
effects that may, or may not distort the resistance at low fields, e.g. as
resistive correction due flux-flow. In any event, considering the
perpendicular field orientation one does not expect to change dramatically
our main conclusions beyond some minor downward shifts of the SC\ critical
field, and possible additional linearly increasing field-dependent
resistivity at low field which may be extended to enhancement of the MR
peaks at high fields, as indeed observed experimentally in Ref.\cite%
{GantmakherPhysicaB2000} for the amorphous InO thin film, as well as in \cite%
{Mograbi19} for the SrTiO$_{3}$/LaAlO$_{3}$ (111) interface.

The proposed dual scenario seems also to be consistent with our previous
TDGL analysis \cite{MZPRB2021},\cite{MZJPC2023} of the field-induced SIT
observed experimentally \cite{Mograbi19} in the SrTiO$_{3}$/LaAlO$_{3}$
(111) interface, though the high-field asymptotic QP conductivity component
used in Ref.\cite{MZPRB2021} has been selected on the basis of the negative
MR measured experimentally in the normal state \cite{RoutPRB2017}, which was
also supported theoretically in \cite{DiezPRL2015}.

\begin{figure}[tbh]
\subfloat[
	The specific relevant fitting
	parameters are: $\protect \varepsilon _{SO}=5.6$ meV, $T_{Q}=150,148,146,142,128$ mK (in respective order of increasing temperature). 
	]{\includegraphics[width=0.45\columnwidth]{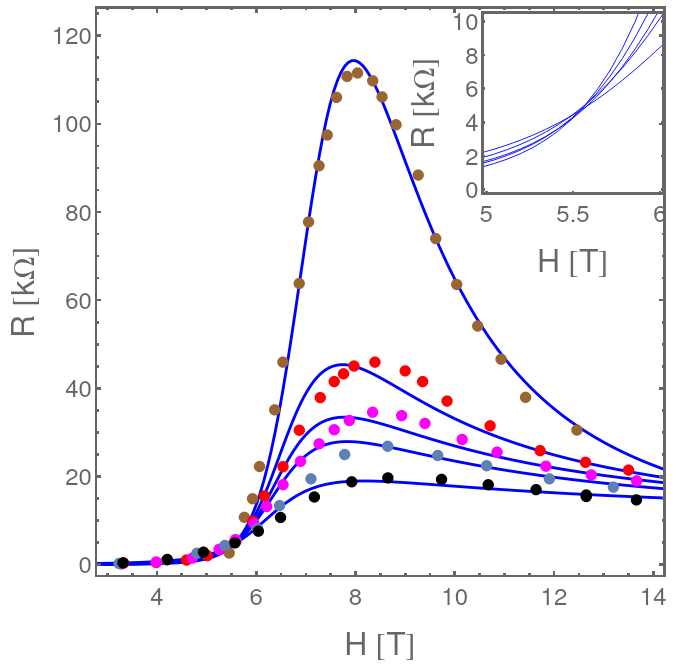}} \hfill 
\subfloat[
	The specific relevant fitting
	parameters are: $\protect \varepsilon _{SO}=5.4$ meV, $T_{Q}=148,142,142,135,125$ mK. 
	]{\includegraphics[width=0.45\columnwidth]{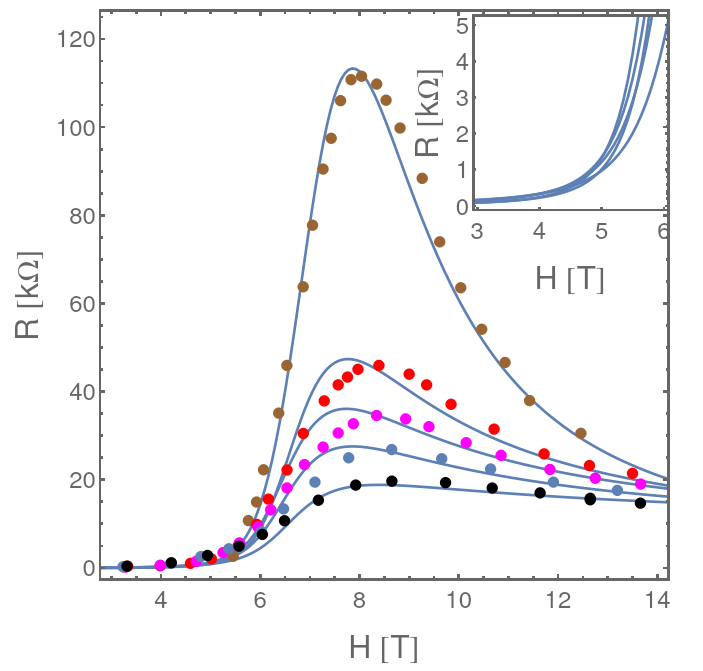}} \newline
\caption{The sheet resistance isotherms calculated at a series of increasing
temperatures; $T=32,60,79,114,195$ mK (their maxima are decreasing
respectively with increasing $T$), plotted together with the corresponding
experimentally measured data (full circles). Insets: Zoom into the crossing
regions of the isotherms showing their dramatic convergence into a single
crossing point by moving from (b) to (a) upon narrowing the distribution of
the quantum tunneling-pair breaking "temperature" $T_{Q}$ over temperature $%
T $. The other parameters used for both (a) and (b) are: $\ T_{c0}=0.8$ K, $%
E_{F}=50$ meV, $\hbar /\protect \tau _{OR}=10$ meV, and: $x_{c0}\equiv \hbar
Dq_{c}^{2}/4\protect \pi k_{B}T_{c0}=0.01$}
\label{fig:fig5}
\end{figure}

\bigskip

\appendix

\bigskip

\section{The reduced stiffness coefficient at low temperatures}

Consider the thermal reduced stiffness parameter $\eta \left( H\right) $ (Eq.%
\ref{eta}) in the low temperature limit, i.e. for $4\pi k_{B}T\ll
\varepsilon _{SO}$, with:%
\begin{equation*}
\delta \equiv \left[ 1-\left( \frac{2\mu _{B}H}{\varepsilon _{SO}}\right)
^{2}\right] ^{1/2}
\end{equation*}

Assuming, first, that the magnitude of the SOS energy parameter $\varepsilon
_{SO}$ is not restricted to large values, we identify:

\begin{equation*}
a_{\pm }=\frac{1}{2}\left( 1\pm \frac{1}{\delta }\right) ,f_{\pm }=\frac{%
\varepsilon _{SO}}{4\pi k_{B}T}\left( 1\pm \delta \right) \gg 1
\end{equation*}%
so that by exploiting the asymptotic form of the digamma-function derivative
we find: 
\begin{eqnarray*}
\eta \left( H\right) \rightarrow a_{+}/f_{-}+a_{-}/f_{+}&=&\left[ \left(
\delta +1\right) /\left( 1-\delta \right) +\left( \delta -1\right) /\left(
1+\delta \right) \right] 2\pi k_{B}T/\varepsilon _{SO}\delta = \\
\\
\ 8\pi k_{B}T/\varepsilon _{SO}\left( 1-\delta ^{2}\right) &=&2\pi
k_{B}T\varepsilon _{SO}/\left( \mu _{B}H\right) ^{2},
\end{eqnarray*}
and:

\begin{equation}
\eta \left( H\right) \rightarrow \frac{2T}{T_{H}},T_{H}\equiv \frac{\left(
\mu _{B}H\right) ^{2}}{\pi k_{B}\varepsilon _{SO}}  \label{eta(H)lowT}
\end{equation}

Now, consider a quantum pair-breaking with the shift $T_{Q}/2T$ introduced
to the reduced stiffness function, Eq.\ref{eta_U}, in the limit $T\ll \left(
T_{Q}+T_{H}\right) /2$ and for sufficiently large SOS energy satisfying $%
\left( 2\mu _{B}H/\varepsilon _{SO}\right) ^{2}\ll 1$:

\begin{equation}
\eta _{U}\left( H\right) \simeq \psi ^{\prime }\left( \frac{1}{2}+\frac{%
T_{Q}+T_{H}}{2T}\right) \rightarrow \frac{2T}{T_{Q}+T_{H}}
\label{eta_U(H)lowT}
\end{equation}

Introducing the tunneling effect, that is:

\begin{equation*}
\Theta \left( H;T_{Q}\right) \equiv \left( 1+\frac{T_{Q}}{T}\right) \eta
_{U}\left( H\right) \rightarrow \left( \frac{T}{T_{Q}}+1\right) \frac{2}{1+%
\frac{T_{H}}{T_{Q}}}
\end{equation*}%
we look for the field $H$\ at which $\Theta \left( H;T_{Q}\right) $
crossovers from decreasing to increasing function of $T_{Q}$, that is at
which: 
\begin{equation*}
\partial \Theta \left( H;T_{Q}\right) /\partial T_{Q}=0.\Longrightarrow
T_{H_{cross}}=T
\end{equation*}%
so that:

\begin{equation}
\Theta \left( H_{cross};T_{Q}\right) =\left( 1+\frac{T_{Q}}{T}\right) \left[
\eta _{U}\left( H_{cross}\right) \right] _{T_{H_{cross}}=T}\rightarrow
\left( \frac{T}{T_{Q}}+1\right) \frac{2}{1+\frac{T}{T_{Q}}}=2  \notag
\end{equation}%
which is independent of $T_{Q}$. The crossover field $H_{cross}$ is
therefore also a crossing point of all curves $\Theta \left( H;T_{Q}\right) $
as functions of $H$, labeled by $T_{Q}$ (see Fig.2). \ Note that for values
of $T_{Q}\leq T_{H}$ the argument of the digamma derivative in Eq.\ref%
{eta_U(H)lowT} at $H\approx H_{cross}$ is of the order unity where its
asymptotic limit is invalid. That is the reason for the significant
deviation of the curve with $T_{Q}=0$ from the crossing point, which is also
not an exact crossing point for finite $T_{Q}$ values, though becoming more
and more accurate by increasing $T_{Q}$.

\bigskip

\section{The CPF momentum distribution function}

Starting with the TDGL-Langevin equation (see \cite{LV05}):

\begin{equation*}
\widehat{L}^{-1}\phi \left( \mathbf{r},t\right) =\zeta \left( \mathbf{r}%
,t\right)
\end{equation*}%
under the white-noise condition of the Langevin force correlator:%
\begin{equation}
\left \langle \zeta ^{\ast }\left( \mathbf{r},t\right) \zeta \left( \mathbf{r%
}^{\prime },0\right) \right \rangle =2k_{B}T\hbar \gamma _{GL}\left(
H\right) \delta \left( \mathbf{r}-\mathbf{r}^{\prime }\right) \delta \left(
t\right)  \label{LFcorrel}
\end{equation}%
the (field-dependent) damping parameter $\gamma _{GL}\left( H\right) \equiv 
\widetilde{\gamma }_{GL}\left( H\right) \pi \alpha /8$ should be determined
self consistently with the life time of the fluctuation modes. Here:%
\begin{equation*}
\alpha =\frac{4\pi ^{2}k_{B}T}{7\zeta \left( 3\right) E_{F}}
\end{equation*}%
and the TDGL propagator $\widehat{L}$ in the frequency-wavenumber
representation is related to the microscopically derived dynamical
fluctuation propagator \cite{MZ-arXiv2025} in the small wavenumber
approximation (see Eq.\ref{FlucProp}) as follows : 
\begin{equation}
\mathcal{D}\left( q,i\Omega _{\nu }\rightarrow \omega \right) \simeq \frac{1%
}{N_{2D}}\frac{1}{\varepsilon _{H}+\eta \left( H\right) \hbar \frac{%
Dq^{2}-i\omega }{4\pi k_{B}T}}=\frac{\alpha k_{B}T}{N_{2D}}L\left( q,\omega
\right)  \label{D(q,-iw)^H}
\end{equation}

Now, considering the correlation function:%
\begin{equation*}
\left \langle \phi ^{\ast }\left( \mathbf{q};t\right) \phi \left( \mathbf{q}%
;0\right) \right \rangle =2k_{B}T\hbar \gamma _{GL}\left( H\right) \int 
\frac{d\omega }{2\pi }e^{-i\omega t}\left \vert L\left( q,\omega \right)
\right \vert ^{2}
\end{equation*}%
and using Eq.\ref{D(q,-iw)^H}, the frequency integration is easily performed
by the residue method, to find:%
\begin{equation}
\left \langle \phi ^{\ast }\left( \mathbf{q};t\right) \phi \left( \mathbf{q}%
;0\right) \right \rangle =\frac{\widetilde{\gamma }_{GL}\left( H\right) }{%
\widetilde{\eta }\left( H\right) }\left( \frac{1}{\alpha }\frac{1}{%
\varepsilon _{H}+\eta \left( H\right) \frac{\hbar Dq^{2}}{4\pi k_{B}T}}%
\right) \exp \left[ -\frac{\varepsilon \left( q;H\right) }{\widetilde{\eta }%
\left( H\right) \gamma _{GL}}t/\hbar \right]  \label{phiphi}
\end{equation}%
where the energy of the $q$ fluctuation-mode is given by:

\begin{equation}
\varepsilon \left( q;H\right) =\alpha \left( k_{B}T\varepsilon _{H}+\frac{%
\eta \left( H\right) \hbar }{4\pi }Dq^{2}\right)  \label{eps(q;H)}
\end{equation}%
and: $\widetilde{\eta }\left( H\right) \equiv \eta \left( H\right) /\eta
\left( 0\right) =2\eta \left( H\right) /\pi ^{2}$.

At this point one note that in order that the (field-dependent) damping
parameter, $\gamma _{GL}\left( H\right) =\widetilde{\gamma }_{GL}\left(
H\right) \gamma _{GL}$, of the Langevin force correlator (Eq.\ref{LFcorrel})
will be determined consistently with the characteristic rate of damping of
the correlation function (see Eq.\ref{phiphi}), $\widetilde{\gamma }%
_{GL}\left( H\right) $ should satisfy the identity:

\begin{equation}
\widetilde{\gamma }_{GL}\left( H\right) =\widetilde{\eta }\left( H\right)
\label{consistCond}
\end{equation}%
which reduces Eq.\ref{phiphi} to the equivalent of the
fluctuation-dissipation theorem, that is:

\begin{equation}
\left \langle \phi ^{\ast }\left( \mathbf{q};t\right) \phi \left( \mathbf{q}%
;0\right) \right \rangle =\left \langle \left \vert \phi \left( \mathbf{q}%
\right) \right \vert ^{2}\right \rangle \exp \left( -\frac{t}{\tau
_{GL}\left( q;H\right) }\right)  \label{FluctDissip}
\end{equation}%
where: 
\begin{equation}
\tau _{GL}\left( q;H\right) \equiv \hbar \frac{\widetilde{\eta }\left(
H\right) \gamma _{GL}}{\varepsilon \left( q;H\right) }=\hbar \frac{\gamma
_{GL}\left( H\right) }{\varepsilon \left( q;H\right) }
\end{equation}%
is the life-time of the fluctuation mode at wavelength $q$. Evidently, as
seen in Eq.\ref{FluctDissip}, the auto-correlation function, $\left \langle
\phi ^{\ast }\left( \mathbf{q};0\right) \phi \left( \mathbf{q};0\right)
\right \rangle $ is found to be equal to the equilibrium momentum
distribution function, $\left \langle \left \vert \phi \left( \mathbf{q}%
\right) \right \vert ^{2}\right \rangle $, given by:

\begin{equation}
\left \langle \left \vert \phi \left( \mathbf{q}\right) \right \vert
^{2}\right \rangle \equiv \frac{1}{\alpha }\frac{1}{\varepsilon _{H}+\eta
\left( H\right) \frac{\hbar Dq^{2}}{4\pi k_{B}T}}  \label{momdistrH}
\end{equation}

It should be emphasized that the validity of the fluctuation-dissipation
theorem, Eq.\ref{FluctDissip}, is bound to the existence of time-reversal
symmetry in the GL hamiltonian, that is to the question whether in Eq.\ref%
{eps(q;H)} $\varepsilon \left( q;-H\right) =\varepsilon \left( q;H\right) $.
The quadratic dependence of $\varepsilon _{H}$ and $\eta \left( H\right) $
on the field $H$, through the field dependence of $a_{\pm }$ and $f_{\pm }$
(see Eq.\ref{af_pm} ), ensures this symmetry.

\bigskip

\section{A model of quantum tunneling out of mesoscopic puddles}

\bigskip

We would like to show that tunneling of CPF out of a mesoscopic puddle can
lead to a recovery of the vanishing stiffness parameter $\eta \left( H\neq
0\right) $ in the zero temperature limit. For this purpose we construct a
simple model hamiltonian of bosons, which tend to aggregate into a 2D
network of puddles, and whose energy dispersion in momentum space follows Eq.%
\ref{eps(q;H)}.

Focusing only on the kinetic part of our toy hamiltonian (i.e. only the part
involved in momentum transfers across the entire network), and assuming in
the first stage of the analysis that the 2D network is a periodic lattice,
it is written in the form: 
\begin{equation}
\mathcal{H}=\sum \limits_{\mathbf{q}}\hbar D\left( H\right) q^{2}\widehat{b}%
_{\mathbf{q}}^{\dagger }\widehat{b}_{\mathbf{q}}+\mathcal{H}_{T}
\label{HamPspace}
\end{equation}%
where $D\left( H\right) \equiv \eta \left( H\right) D/4\pi $, and: 
\begin{equation*}
\mathcal{H}_{T}=\sum \limits_{\mathbf{q,q}^{\prime }}U_{\mathbf{q,q}^{\prime
}}\widehat{b}_{\mathbf{q}}^{\dagger }\widehat{b}_{\mathbf{q}^{\prime }}
\end{equation*}%
is the tunneling hamiltonian, $\widehat{b}_{\mathbf{q}}^{\dagger },\widehat{b%
}_{\mathbf{q}}$ are creation and annihilation operators respectively of a
boson with momentum $\mathbf{q}$, and:

\begin{equation*}
U_{\mathbf{q,q}^{\prime }}=\int d^{2}r\chi _{\mathbf{q}}^{\ast }\left( 
\mathbf{r}\right) U\left( \mathbf{r}\right) \chi _{\mathbf{q}^{\prime
}}\left( \mathbf{r}\right)
\end{equation*}
the tunneling matrix elements between Bloch functions:%
\begin{equation*}
\chi _{\mathbf{q}}\left( \mathbf{r}\right) =\frac{1}{\sqrt{N}}\sum
\limits_{j=1}^{N}e^{i\mathbf{q\cdot R}_{j}}\chi \left( \mathbf{r-R}%
_{j}\right)
\end{equation*}%
with $\chi \left( \mathbf{r-R}_{j}\right) $ a (Wannier) wave function of a
boson in a puddle localized around $\mathbf{R}_{j}$. In analogy with the
tight-binding approach, $U\left( \mathbf{r}\right) <0$ represents in our
model the interaction potential between a boson and the underlying ionic
lattice.

In real space the toy boson hamiltonian reads:

\begin{equation}
\mathcal{H}=\int d^{2}r\hbar D\left( H\right) \widehat{\varphi }^{\dagger
}\left( \mathbf{r}\right) \left( -\nabla ^{2}\right) \widehat{\varphi }%
\left( \mathbf{r}\right) +\mathcal{H}_{T}  \label{HamRspace}
\end{equation}%
where:

\begin{eqnarray*}
\widehat{\varphi }\left( \mathbf{r}\right) &=&\sum \limits_{\mathbf{q}}\chi
_{\mathbf{q}}\left( \mathbf{r}\right) \widehat{b}_{\mathbf{q}}, \\
\mathcal{H}_{T} &=&\int d^{2}rU\left( \mathbf{r}\right) \widehat{\varphi }%
^{\dagger }\left( \mathbf{r}\right) \widehat{\varphi }\left( \mathbf{r}%
\right)
\end{eqnarray*}

Averaging the tunneling hamiltonian in Eq.\ref{HamRspace} over the
fluctuations ensemble we have:

\begin{equation*}
\left \langle \mathcal{H}_{T}\right \rangle \approx \sum \limits_{\mathbf{q}%
}n_{\mathbf{q}}\overline{U}+\sum \limits_{\mathbf{q}}n_{\mathbf{q}}\sum
\limits_{NN\left( \mathbf{R}_{jj^{\prime }}\right) }e^{i\mathbf{q\cdot R}%
_{jj^{\prime }}}\int d^{2}r\chi ^{\ast }\left( \mathbf{r}\right) U\left( 
\mathbf{r}\right) \chi \left( \mathbf{r-R}_{jj^{\prime }}\right)
\end{equation*}%
where: $n_{\mathbf{q}}\equiv \left \langle \widehat{b}_{\mathbf{q}}^{\dagger
}\widehat{b}_{\mathbf{q}}\right \rangle $, $\overline{U}\equiv \int
d^{2}r\chi ^{\ast }\left( \mathbf{r}\right) U\left( \mathbf{r}\right) \chi
\left( \mathbf{r}\right) $, and $\mathbf{R}_{jj^{\prime }}\equiv \mathbf{R}%
_{j^{\prime }}-\mathbf{R}_{j}$ are restricted to $\mathbf{R}_{j^{\prime }}$
nearest neighbors (NN) to $\mathbf{R}_{j}$.

Assuming some disorder in the positions $\mathbf{R}_{j^{\prime }}$ and
averaging over all their possible configurations, we get for the tunneling
energy:

\begin{equation*}
\left \langle \left \langle \mathcal{H}_{T}\right \rangle \right \rangle
\approx \sum \limits_{\mathbf{q}}n_{\mathbf{q}}\overline{U}-\sum \limits_{%
\mathbf{q}}q^{2}n_{\mathbf{q}}\sum \limits_{NN\left( \mathbf{R}_{jj^{\prime
}}\right) }\left \langle R_{jj^{\prime }}^{2}\cos ^{2}\theta _{jj^{\prime
}}U_{jj^{\prime }}\right \rangle
\end{equation*}

\begin{equation*}
U_{jj^{\prime }}\equiv \int d^{2}r\chi ^{\ast }\left( \mathbf{r}\right)
U\left( \mathbf{r}\right) \chi \left( \mathbf{r-R}_{jj^{\prime }}\right)
\end{equation*}%
where the second average symbol corresponds to averaging over NN positions.

Defining a characteristic average NN distance $\zeta _{0}$ and NN matrix
elelment $U_{NN}$, respectively by:

\begin{eqnarray*}
\zeta _{0}^{2} &\equiv &\left \langle R_{jj^{\prime }}^{2}\cos ^{2}\theta
_{jj^{\prime }}\right \rangle , \\
U_{NN} &\equiv &-\frac{1}{\zeta _{0}^{2}}\sum \limits_{NN\left( \mathbf{R}%
_{jj^{\prime }}\right) }\left \langle R_{jj^{\prime }}^{2}\cos ^{2}\theta
_{jj^{\prime }}U_{jj^{\prime }}\right \rangle
\end{eqnarray*}%
we get for the boson energy dispersion term, to second order in small $q$: 
\begin{equation}
\left \langle \left \langle \mathcal{H}\right \rangle \right \rangle
\rightarrow \sum \limits_{\mathbf{q}}\left[ \eta \left( H\right) \frac{\hbar
D}{4\pi }+\zeta _{0}^{2}U_{NN}\right] q^{2}n_{\mathbf{q}}  \label{aveHdisp}
\end{equation}

It is evident that the total reduced stiffness coefficient in Eq.\ref%
{aveHdisp}; $\eta \left( H\right) +4\pi \zeta _{0}^{2}U_{NN}/\hbar D$
remains finite, equal to: $\Delta \eta _{tunn}=4\pi \zeta
_{0}^{2}U_{NN}/\hbar D$, in the zero temperature limit for any $H\neq 0$.

\bigskip

\section{Phenomenological approach to Tunneling-Pair-breaking of CPFs bosons}

\bigskip

We start from the CPFs partition function in the Hartree approximation \cite%
{MZPRB2021}:

\begin{equation}
Z_{fluct}\rightarrow \prod \limits_{\mathbf{q}}\int \mathcal{D}\left( \left
\vert \Delta \left( q\right) \right \vert ^{2}\right) \exp \left \{ -\frac{%
\tau _{T}}{\hbar }\left[ \varepsilon _{H}+\left( \frac{\tau _{T}}{\hbar }%
\right) ^{-1}\sum \limits_{n=0}^{\infty }\left( 
\begin{array}{c}
\pi f_{2}\left( \hbar \omega _{n};H\right) \hbar Dq^{2}+ \\ 
8f_{4}\left( \hbar \omega _{n};H\right) \int \frac{d^{2}q^{\prime }}{\left(
2\pi \right) ^{2}}\mathcal{D}_{T}\left( q^{\prime }\right)%
\end{array}%
\right) \right] \left \vert \Delta \left( q\right) \right \vert ^{2}\right \}
\label{Z_f}
\end{equation}%
where the thermal time interval is: $\tau _{T}\equiv \hbar /k_{B}T$, the
thermal fluctuation propagator: $\mathcal{D}_{T}\left( q\right) \equiv \left
\langle \left \vert \Delta \left( q\right) \right \vert ^{2}\right \rangle
=k_{B}T\mathcal{D}\left( q\right) $, and the Matsubara frequency-dependent
two and four electron correlation functions are defined by:

\begin{eqnarray}
f_{2}\left( \hbar \omega _{n};H\right) &\equiv &\frac{\left[ \left( \hbar
\omega _{n}+\varepsilon _{SO}\right) ^{2}-\left( \mu _{B}H\right) ^{2}\right]
}{\left[ \hbar \omega _{n}\left( \hbar \omega _{n}+\varepsilon _{SO}\right)
+\left( \mu _{B}H\right) ^{2}\right] ^{2}}=\boldsymbol{\mathbf{Re}}\left[ 
\frac{\hbar \omega _{n}+\varepsilon _{SO}+i\mu _{B}H}{\hbar \omega
_{n}\left( \hbar \omega _{n}+\varepsilon _{SO}\right) +\left( \mu
_{B}H\right) ^{2}}\right] ^{2},  \label{f2} \\
f_{4}\left( \hbar \omega _{n};H\right) &\equiv &\frac{\left( \hbar \omega
_{n}+\varepsilon _{SO}\right) \left[ \left( \hbar \omega _{n}+\varepsilon
_{SO}\right) ^{2}-3\left( \mu _{B}H\right) ^{2}\right] }{\left[ \hbar \omega
_{n}\left( \hbar \omega _{n}+\varepsilon _{SO}\right) +\left( \mu
_{B}H\right) ^{2}\right] ^{3}}=\boldsymbol{\mathbf{Re}}\left[ \frac{\hbar
\omega _{n}+\varepsilon _{SO}+i\mu _{B}H}{\hbar \omega _{n}\left( \hbar
\omega _{n}+\varepsilon _{SO}\right) +\left( \mu _{B}H\right) ^{2}}\right]
^{3}  \label{f4}
\end{eqnarray}%
with the fermionic Matsubara frequency: $\omega _{n}=\left( 2n+1\right) \pi
k_{B}T/\hbar $.

In the corresponding (static) fluctuation propagator $\mathcal{D}\left(
q\right) $:

\begin{equation}
\left[ N_{2D}\mathcal{D}\left( q\right) \right] ^{-1}=\varepsilon
_{H}+\left( \frac{\tau _{T}}{\hbar }\right) ^{-1}\sum \limits_{n=0}^{\infty
}\left( 
\begin{array}{c}
\pi f_{2}\left( \hbar \omega _{n};H\right) \hbar Dq^{2}+ \\ 
8f_{4}\left( \hbar \omega _{n};H\right) \int \frac{d^{2}q^{\prime }}{\left(
2\pi \right) ^{2}}\mathcal{D}_{T}\left( q^{\prime }\right) 
\end{array}%
\right)   \notag
\end{equation}%
we identify the reduced stiffness $\eta \left( H\right) $ and the
interaction $\mathcal{F}\left( H\right) $ functions in: $\sum \limits_{n=0}^{%
\infty }f_{2}\left( \hbar \omega _{n};H\right) =\eta \left( H\right) /\left(
2\pi k_{B}T\right) ^{2}$, and: $\sum \limits_{n=0}^{\infty }f_{4}\left( \hbar
\omega _{n};H\right) =\mathcal{F}\left( H\right) /\left( 2\pi k_{B}T\right)
^{3}$, respectively.

The first step of implementing the joint quantum tunneling-pair-breaking
effects within our phenomenological approach is to introduce quantum
tunneling into the partition function by replacing the thermal time interval 
$\tau _{T}\equiv \hbar /k_{B}T$, appearing in Eq.\ref{Z_f}, with the unified
quantum-thermal time interval $\tau _{U}$, according to:

\begin{equation*}
\frac{1}{\tau _{T}}\rightarrow \frac{1}{\tau _{U}}=\frac{1}{\tau _{T}}+\frac{%
1}{\tau _{Q}}\equiv k_{B}\left( T+T_{Q}\right) /\hbar
\end{equation*}%
where $\tau _{Q}\equiv \hbar /k_{B}T_{Q}$, so that the corresponding
effective ("dressed") Gaussian fluctuation propagator is obtained by the
replacement:

\begin{equation}
\mathcal{D}_{T}\left( q\right) ^{-1}\rightarrow \mathcal{D}_{u}\left(
q\right) ^{-1}=\frac{N_{2D}}{k_{B}\left( T+T_{Q}\right) }\left[ \varepsilon
_{H}+k_{B}\left( T+T_{Q}\right) \sum \limits_{n=0}^{\infty }\left( 
\begin{array}{c}
\pi f_{2}\left( \hbar \omega _{n};H\right) \hbar Dq^{2}+ \\ 
8f_{4}\left( \hbar \omega _{n};H\right) \int \frac{d^{2}q^{\prime }}{\left(
2\pi \right) ^{2}}\mathcal{D}_{u}\left( q^{\prime }\right)%
\end{array}%
\right) \right]  \label{D_u}
\end{equation}

This replacement is equivalent to the modification introduced to the
stiffness function due to tunneling: 
\begin{equation*}
\eta \left( H\right) \rightarrow \left( 1+T_{Q}/T\right) \eta \left(
H\right) \equiv \eta \left( H\right) +\Delta \eta _{tunn}
\end{equation*}%
as can be seen by considering the two-electron term in Eq.\ref{D_u}, i.e.: 
\begin{equation*}
k_{B}\left( T+T_{Q}\right) \sum \limits_{n=0}^{\infty }\pi f_{2}\left( \hbar
\omega _{n};H\right) \hbar Dq^{2}=\left[ \left( 1+T_{Q}/T\right) \eta \left(
H\right) /4\pi k_{B}T\right] \hbar Dq^{2}=\left[ \left( \eta \left( H\right)
+\Delta \eta _{tunn}\right) /4\pi k_{B}T\right] \hbar Dq^{2},
\end{equation*}%
which yields for the "bare" Gaussian propagator:

\begin{equation*}
\mathcal{D}_{T}^{Gauss}\left( q\right) \rightarrow \mathcal{D}%
_{u}^{Gauss}\left( q\right) =\frac{k_{B}\left( T+T_{Q}\right) }{N_{2D}\left[
\varepsilon _{H}+\left( 1+T_{Q}/T\right) \eta \left( H\right) \frac{\hbar
Dq^{2}}{4\pi k_{B}T}\right] }
\end{equation*}

Note the use of the lower case subscript u which indicates the initial
implementation of tunneling without pair breaking (see below).

\bigskip

For the interaction term this replacement yields: 
\begin{eqnarray*}
8k_{B}\left( T+T_{Q}\right) \sum \limits_{n=0}^{\infty }f_{4}\left( \hbar
\omega _{n};H\right) \int \frac{d^{2}q^{\prime }}{\left( 2\pi \right) ^{2}}%
\mathcal{D}_{u}\left( q^{\prime }\right)  &=&8k_{B}\left( T+T_{Q}\right) 
\frac{\mathcal{F}\left( H\right) }{\left( 2\pi k_{B}T\right) ^{3}}\int \frac{%
d^{2}q^{\prime }}{\left( 2\pi \right) ^{2}}\frac{k_{B}\left( T+T_{Q}\right) 
}{N_{2D}\left[ \varepsilon _{H}+\frac{\eta \left( H\right) +\Delta \eta
_{tunn}}{4\pi k_{B}T}\hbar Dq^{\prime 2}\right] } \\
&=&\frac{2\left( 1+T_{Q}/T\right) \mathcal{F}\left( H\right) /\eta \left(
H\right) }{\pi ^{2}E_{F}\tau /\hbar }\ln \left( 1+\frac{\zeta _{c}\left(
H\right) }{\varepsilon _{H}}\right) ,
\end{eqnarray*}%
so that finally:

\begin{eqnarray}
Z_{fluct}^{u} &\rightarrow &\prod \limits_{\mathbf{q}}\int \mathcal{D}\left(
\left \vert \Delta \left( q\right) \right \vert ^{2}\right) \times
\label{Z_fl^u} \\
&&\exp \left \{ -\frac{\tau _{U}}{\hbar }\left[ \varepsilon _{H}+\left(
1+T_{Q}/T\right) \left( 
\begin{array}{c}
\frac{\eta \left( H\right) }{4\pi k_{B}T}\hbar Dq^{2}+ \\ 
\frac{2}{\pi ^{2}}\frac{1}{E_{F}\tau /\hbar }\frac{\mathcal{F}\left(
H\right) }{\eta \left( H\right) }\ln \left( 1+\frac{x_{c}}{\varepsilon _{H}}%
\right)%
\end{array}%
\right) \right] \left \vert \Delta \left( q\right) \right \vert ^{2}\right \}
\notag
\end{eqnarray}

In order to sustain the dynamical equilibrium between the 2D system of CPFs'
bosons and the unpaired normal-state electron gas within the process of
implementing quantum tunneling effects, one should introduce the effect of
quantum pair-breaking that consistently follows the tunneling of CPFs out of
the mesoscopic puddles. This is done through the Matsubara
frequency-dependent two-electron and four-electron correlation functions $%
f_{2}\left( \hbar \omega _{n};H\right) $,and $f_{4}\left( \hbar \omega
_{n};H\right) $, Eqs.\ref{f2} and \ref{f4} \ respectively, controlling the
partition function, Eq.\ref{Z_f}, and the propagator, Eq.\ref{D_u}.
Consistently with this, the critical shift function $\varepsilon _{H}$
should be also appropriately modified.

Thus, the thermal equilibrium many-electron correlation functions $%
f_{2}\left( \hbar \omega _{n};H\right) $, $f_{4}\left( \hbar \omega
_{n};H\right) $,.... are modified to take into account "excited" states
associated with the tunneling operator: $\left( 1+T_{Q}/T\right) \eta \left(
H\right) =\eta \left( H\right) +\Delta \eta _{tunn}$, by shifting, under
summation, the equilibrium Matsubara frequency $\omega _{n}$ to the
"excitation" frequency: $\omega _{n}+\pi T_{Q}/\hbar =2\pi k_{B}T\left[
n+\left( 1/2\right) \left( 1+T_{Q}/T\right) \right] /\hbar $, and define the
unified, quantum-thermal (QT) correlation functions:

\begin{eqnarray}
\eta _{U}\left( H\right) &=&\left( 2\pi k_{B}T\right) ^{2}\sum
\limits_{n=0}^{\infty }f_{2}\left( \hbar \omega _{n}+\pi T_{Q};H\right) ,
\label{eta_U(H)Mats} \\
\mathcal{F}_{U}\left( H\right) &=&\left( 2\pi k_{B}T\right) ^{3}\sum
\limits_{n=0}^{\infty }f_{4}\left( \hbar \omega _{n}+\pi T_{Q};H\right)
\label{F_U(H)}
\end{eqnarray}

A similar "excitation" imaginary frequency shift is introduced to the
critical shift parameter $\varepsilon _{H}$ (see Eq.\ref{eps_H}), through
the digamma functions, which transforms to: 
\begin{equation}
\varepsilon _{H}^{U}\equiv \ln \left( \frac{T}{T_{c0}}\right) +a_{+}\psi %
\left[ \frac{1}{2}\left( 1+\frac{T_{Q}}{T}\right) +f_{-}\right] +a_{-}\psi %
\left[ \frac{1}{2}\left( 1+\frac{T_{Q}}{T}\right) +f_{+}\right] -\psi \left( 
\frac{1}{2}\right)  \label{eps^U_H}
\end{equation}%
so that, finally, the corresponding effective ("dressed") Gaussian
propagator, obtained from Eq.\ref{D_u}, is:

\begin{equation}
\mathcal{D}_{U}\left( q\right) ^{-1}=\frac{N_{2D}}{k_{B}\left(
T+T_{Q}\right) }\left[ \varepsilon _{H}+\left( 1+\frac{T_{Q}}{T}\right) \left(%
\begin{array}{c}
\frac{\eta _{U}\left( H\right) }{4k_{B}T}%
\hbar Dq^{2}\hbar Dq^{2}+ \\ 
\frac{2}{\pi ^{2}} \frac{\mathcal{F}%
_{U}\left( H\right) }{\left( k_{B}T\right) ^{2}}\int \frac{d^{2}q^{\prime }}{%
\left( 2\pi \right) ^{2}}\mathcal{D}_{U}\left( q^{\prime }\right)%
\end{array}%
\right)\right]  \label{IntEqD}
\end{equation}

Eq.\ref{IntEqD} may be considered as an integral equation for the propagator 
$\mathcal{D}_{U}\left( q\right) $ in the Hartree approximation. Since the
interaction term is independent of $q$ one may redefine the critical shift
parameter as:

\begin{equation}
\widetilde{\varepsilon }_{H}^{U}\equiv \varepsilon _{H}^{U}+\left( 1+\frac{%
T_{Q}}{T}\right) \frac{2}{\pi ^{2}}\frac{\mathcal{F}_{U}\left( H\right) }{%
\left( k_{B}T\right) ^{2}}\int \frac{d^{2}q^{\prime }}{\left( 2\pi \right)
^{2}}\mathcal{D}_{U}\left( q^{\prime };\widetilde{\varepsilon }%
_{H}^{U}\right)  \label{SCeq}
\end{equation}%
in which considering the propagator under integration as a function of the
"dressed" critical shift parameter $\widetilde{\varepsilon }_{H}^{U}$ leads
to equation for the latter equivalent to the integral equation \ref{IntEqD}.

Thus, performing the integration over $q$ in Eq.\ref{SCeq} the resulting
self-consistent-field (SCF) equation for $\widetilde{\varepsilon }_{H}^{U}$,
i.e.:

\begin{equation}
\widetilde{\varepsilon }_{H}^{U}=\varepsilon _{H}^{U}+\frac{2}{\pi ^{2}}%
\frac{1}{E_{F}\tau /\hbar }\left( 1+\frac{T_{Q}}{T}\right) \frac{\mathcal{F}%
_{U}\left( H\right) }{\eta _{U}\left( H\right) }\ln \left( 1+\frac{\zeta
_{c}\left( H\right) }{\widetilde{\varepsilon }_{H}^{U}}\right)
\label{SCFeq^U}
\end{equation}%
provides solutions to the integral equation \ref{IntEqD} for the propagator $%
\mathcal{D}_{U}\left( q\right) $.

\bigskip

\section{Invariance of the cutoff for the UV divergence}

\bigskip

We start with the integral over $q^{2}$ in Eq.\ref{n_S(H)} for the CPFs
density, 
\begin{equation*}
n_{CPF}\left( H\right) =\left( \frac{7\zeta \left( 3\right) E_{F}}{4\pi
^{2}\hbar D}\right) \frac{1}{d}I\left( H\right)
\end{equation*}

which can be transformed into an integral over the dimensionless variable $%
x\equiv \hbar Dq^{2}/4\pi k_{B}T$:

\begin{equation}
I\left( H\right) =\int_{0}^{x_{c}}\frac{dx}{\varepsilon _{H}+\eta \left(
H\right) x}=\frac{1}{\eta \left( H\right) }\int_{0}^{\zeta _{c}\left(
H\right) }\frac{d\zeta }{\varepsilon _{H}+\zeta }=\frac{1}{\eta \left(
H\right) }\ln \left( 1+\frac{\zeta _{c}\left( H\right) }{\varepsilon _{H}}%
\right)  \label{I(H)}
\end{equation}%
where the field-dependent dimensionless cutoff is:

\begin{equation*}
\zeta _{c}\left( H\right) \equiv \eta \left( H\right) \frac{\hbar D}{4\pi
k_{B}T}q_{c}^{2}
\end{equation*}

In the presence of quantum tunneling the reduced stiffness undergoes a
shift: $\eta _{+}\left( H\right) \equiv \eta \left( H\right) +\Delta \eta
_{tunn}$, but the corresponding integral over $x$ is rewritten, similar to
the integral in Eq.\ref{I(H)}, as a universal integral over $\zeta $ whose
UV divergence determines a cutoff identical to $\zeta _{c}\left( H\right) $,
so that:

\begin{equation*}
I_{+}\left( H\right) =\int_{0}^{x_{c}}\frac{dx}{\varepsilon _{H}+\eta
_{+}\left( H\right) x}=\frac{1}{\eta _{+}\left( H\right) }\int_{0}^{\zeta
_{c}\left( H\right) }\frac{d\zeta }{\varepsilon _{H}+\zeta }=\frac{1}{\eta
_{+}\left( H\right) }\ln \left( 1+\frac{\zeta _{c}\left( H\right) }{%
\varepsilon _{H}}\right)
\end{equation*}


\bigskip

\begin{thebibliography}{99}
\bibitem{Clogston1962} A. M. Clogston, "Upper Limit for the Critical Field
in Hard Superconductors", Phys. Rev. Lett., \textbf{9}, 266 (1962).

\bibitem{Chandrasekhar1962} B. S. Chandrasekhar, "A Note on the Maximum
Critical Field of High-Field Superconductors" Appl. Phys. Lett., \textbf{1},
7 (1962).

\bibitem{AdamsRev2012} P. W. Adams, "Spin Effects Near the
Superconductor--Insulator Transition" in Conductor-Insulator Quantum Phase
Transitions, edited: V. Dobrosavljevic, N. Trivedi, J. M. Valles, Jr.,
(Oxford Scholarship Online, 2012).

\bibitem{AdamsPRL94} Wenhao Wu and P.W. Adams, "Superconductor-Insulator
Transition in a Parallel Magnetic Field", Phys. Rev. Lett. \textbf{73}, 1412
(2012).

\bibitem{Fisher90} M. P. A. Fisher, "Quantum Phase Transitions in Disordered
Two-Dimensional Superconductors", Phys. Rev. Lett. \textbf{65}, 923 (1990).

\bibitem{GantmakherPhysicaB2000} V.F. Gantmakher, M.V. Golubkov, V.T.
Dolgopolov, G.E. Tsydynzhapov, A.A. Shashkin, "Superconductor-insulator
transition in amorphous In-O films", Physica B 284-288, 649-650 (2000)

\bibitem{GantmakherJETPLett2000} V. F. Gantmakher, M. V. Golubkov, V. T.
Dolgopolov, A. A. Shashkin, and G. E. Tsydynzhapov, "Observation of the
Parallel-Magnetic-Field-Induced Superconductor--Insulator Transition in Thin
Amorphous InO Films", JETP Letters, \textbf{71} 11, 473--476 (2000).

\bibitem{Mograbi19} M. Mograbi, E. Maniv, P. K. Rout, D. Graf, J. -H Park
and Y. Dagan, "Vortex excitations in the Insulating

State of an Oxide Interface", Phys. Rev. B 99, 094507 (2019).

\bibitem{RoutPRL2017} P. K. Rout, E. Maniv, and Y. Dagan, "Link between the
Superconducting Dome and Spin-Orbit Interaction in the (111) LaAlO$_{3}$%
/SrTiO$_{3}$ Interface", Phys. Rev. Lett. 119, 237002 (2017).

\bibitem{ManivNatCommun2015} E. Maniv, M. Ben Shalom, A. Ron, M. Mograbi, A.
Palevski, M. Goldstein and Y. Dagan, "Strong correlations elucidate the
electronic structure and phase diagram of LaAlO$_{3}$/SrTiO$_{3}$
interface", Nat. Commun. \textbf{6}, 8239 (2015).

\bibitem{Ohtomo04} A. Ohtomo, and H. Y. Hwang, "A high-mobility electron gas
at the LaAlO$_{3}$/SrTiO$_{3}$ heterointerface", Nature 427, 423 (2004).

\bibitem{Caviglia08} A. D. Caviglia, S. Gariglio, N. Reyren, D. Jaccard, T.
Schneider, M. Gabay, S. Thiel, G. Hammerl, J. Mannhart and J.-M. Triscone,
"Electric field control of the LaAlO3/SrTiO3 interface ground state", Nature
(London) 456, 624 (2008).

\bibitem{GantmakherJETP96} V. F. Gantmakher and M. V. Golubkov, J. G. S. Lok
and A. K. Geim, "Giant negative magnetoresistance of semi-insulating
amorphous indium oxide films in strong magnetic fields",
Zh. Eksp. Teor. Fiz. \textbf{109}, 1765 (1996)[Sov. Phys. JETP \textbf{82}, 951 (1996)].

\bibitem{Fukuyama81} S. Maekwa and H. Fukuyama, "MR in 2D disordered
systems: Effects of Zeeman splitting and SO scattering", J. Phys. Soc. Jpn. 
\textbf{50, }2516 (1981).

\bibitem{Fukuyama1985} H. Fukuyama, "Interaction effects in the weakly
localized regime of two-and three-dimensional disordered systems", in Modern
Problems in Condensed Matter Sciences, volume 10, ed. A.L. Efros and M.
Pollak, Northholland 1985.

\bibitem{DiezPRL2015} M. Diez, A. M. R. V. L. Monteiro, G. Mattoni, E.
Cobanera, T. Hyart, E. Mulazimoglu, N. Bovenzi, C. W. J. Beenakker, and A.
D. Caviglia, "Giant Negative Magnetoresistance Driven by Spin-Orbit Coupling
at the LaAlO$_{3}$/SrTiO$_{3}$ Interface", Phys. Rev. Lett. \textbf{115},
016803 (2015).

\bibitem{MZPRB2021} T. Maniv and V. Zhuravlev, "Superconducting fluctuations
and giant negative magnetoresistance in a gate-voltage tuned \
two-dimensional electron system with strong spin-orbit impurity scattering",
Phys. Rev. B 104, 054503 (2021).

\bibitem{MZJPC2023} T. Maniv and V. Zhuravlev, \textquotedblleft
Field-induced boson insulating states in a 2D superconducting electron gas
with strong spin--orbit scatterings\textquotedblright , J. Phys.: Condensed
Matter \textbf{35} 055001 (2023).

\bibitem{LV05} A. Larkin and A. Varlamov, "Theory of fluctuations in
superconductors", (Oxford Scholarship Online, 2005 ).

\bibitem{GalitLarkinPRB01} V. M. Galitski and A. I. Larkin, "Superconducting
fluctuations at low temperature", Phys. Rev. B 63, 174506 (2001).

\bibitem{Glatzetal2011} A. Glatz, A. A. Varlamov, and V. M. Vinokur,
"Fluctuation spectroscopy of disordered two-dimensional superconductors",
Phys. Rev. B 84, 104510 (2011).

\bibitem{Lopatinetal05} A. V. Lopatin, N. Shah, and V. M. Vinokur,
"Fluctuation Conductivity of Thin Films and Nanowires Near a
Parallel-Field-Tuned Superconducting Quantum Phase Transition", Phys. Rev.
Lett. \textbf{94}, 037003 (2005).

\bibitem{Khodas2012} M. Khodas, A. Levchenko, and G. Catelani,
"Quantum-Fluctuation Effects in the Transport Properties of Ultrathin Films
of Disordered Superconductors above the Paramagnetic Limit", Phys. Rev.
Lett. \textbf{108}, 257004 (2012).

\bibitem{MZ-arXiv2025} T. Maniv and V. Zhuravlev, "Microscopic Transport
Theory of Cooper-pair fluctuations in a disordered 2D Electron System with
Spin-Orbit Scatterings", arXiv:2403.02117 [cond-mat.supr-con]

\bibitem{AL68} L. G. Aslamazov and A.I. Larkin, Phys. Lett. A 26 p. 238
(1968).

\bibitem{AdkinsJPC1980} C J Adkins, J M D Thomas and M W Young, "Increased
resistance below the superconducting transition in granular metals", J.
Phys. C: Solid St. Phys. \textbf{13}, 3427 (1980).

\bibitem{GantmakherJETPLett95} V. F. Gantmakher and M. V. Golubkov,
"Superconductivity and negative MR in amorphous In$_{2}$O$_{x}$ films",Pis’ma Zh. Eksp. Teor. Fiz.\textbf{61}, 593 (1995) [Sov. Phys. JETP Lett. \textbf{61}, 606 (1995)].

\bibitem{Abrikosov62} A. A. Abrikosov and L.P. Gorkov, "Spin-orbit interaction and the knight shift in superconductors" J. Exptl. Theoret. Phys. (U.S.S.R.)
\textbf{42}, 1088 (1962) [Sov. Phys.-JETP \textbf{15}, 752 (1962)]

\bibitem{Klemm75} R. A. Klemm, A. Luther and M.R. Beasley, "Theory of the
upper critical field in layered superconductors", Phys. Rev. B \textbf{12},
877 (1975).\ 

\bibitem{IsingSC} Chong Wang, Yong Xu, and Wenhui Duan,
"Ising Superconductivity and Its Hidden Variants", Acc. Mater. Res. \textbf{2}%
, 526-533 (2021).

\bibitem{HaoranAccMatRes2024} Haoran Ji, Yi Liu , Chengcheng Ji, and Jian
Wang, "Two-Dimensional and Interface Superconductivity in Crystalline
Systems", Acc. Mater. Res. \textbf{5} 1146-1157 (2024).

\bibitem{Maki66} K. Maki, "Effect of Pauli Paramagnetism on Magnetic
Properties of High-Field Superconductors", Phys. Rev. \textbf{148}, 362
(1966).

\bibitem{FuldeMaki70} P. Fulde and K. Maki, "Fluctuations in High Field
Superconductors", Z. Physik \textbf{238}, 233--248 (1970).

\bibitem{WHH66} N. R. Werthamer, E. Helfand, and P. C. Hohenberg,
"Temperature and Purity Dependence of the Superconducting Critical Field H$%
_{c2}$. III. Electron Spin and Spin-Orbit Effects", Phys. Rev. \textbf{147},
295 (1966).

\bibitem{UllDor90} S. Ullah and A.T. Dorsey, "Critical Fluctuations in
High-Temperature Superconductors and the Ettingshausen Effect", Phys. Rev.
Lett. 65, 2066 (1990). Properties of (111).

\bibitem{UllDor91} S. Ullah and A.T. Dorsey, "Effect of fluctuations on the
transport properties of type-II superconductors in a magnetic field", Phys.
Rev. B \textbf{44}, 262 (1991).

\bibitem{DynesPRL1978} R. C. Dynes, J. P. Garno, and J. M. Rowell,
"Two-Dimensional Electrical Conductivity in Quench-Condensed Metal Films",
Phys. Rev. Lett. \textbf{40}, 479 (1978).

\bibitem{GerberPRL1997} A. Gerber, A. Milner, G. Deutscher, M. Karpovsky,
and A. Gladkikh, "Insulator-Superconductor Transition in 3D Granular Al-Ge
Films", Phys. Rev. Lett. \textbf{78}, 4277 (1997).

\bibitem{BeloborodovPRL1999} I. S. Beloborodov and K.B. Efetov, "Negative
Magnetoresistance of Granular Metals in a Strong Magnetic Field", Phys. Rev.
Lett. \textbf{82}, 4277 (1999).

\bibitem{BeloborodovPRB2000} I. S. Beloborodov, K. B. Efetov, A. I. Larkin,
"Magnetoresistance of granular superconducting metals in a strong magnetic
field", Phys. Rev. B \textbf{61}, 9145 (2000).

\bibitem{ManivPRB2026} T. Maniv and V. Zhuravlev, “Field-induced superconductor-insulator transition in disordered two-dimensional electron systems: The case of amorphous indium-oxide thin films”, Phys. Rev. B \textbf{113}, 054503 (2026). 

\bibitem{YeomJMC2016} H.-I. Yeom, J. B. Ko, G. Mun and S.-H. Ko Park, "High
mobility polycrystalline indium oxide thin-film transistors by means of
plasma-enhanced atomic layer deposition", J.Mater. Chem. C 4, 6873 (2016).

\bibitem{ShapirPRB1989} Y. Shapir and Z. Ovadyahu, "Effects of spin-orbit
scattering on hopping magnetoconductivity", Phys. Rev. B \textbf{40}, 12441
(1989).

\bibitem{RoutPRB2017} P. K. Rout, I. Agireen, E. Maniv, M. Goldstein, and Y.
Dagan, Six-fold crystalline anisotropic magnetoresistance in the (111) LaAlO$%
_{3}$/SrTiO$_{3}$ oxide interface, Phys. Rev. B \textbf{95}, 241107(R)
(2017).
\end{thebibliography}
\end{document}